\documentclass[iop]{emulateapj}
\usepackage {graphicx}

\begin{document}

\submitted{Accepted to ApJ, March 14, 2015}

\title{Ages of Star Clusters in the Tidal Tails of Merging Galaxies}

\author{A. J. Mulia\altaffilmark{1},
R. Chandar\altaffilmark{1},
B. C. Whitmore\altaffilmark{2}}

\altaffiltext{1}{Physics \& Astronomy Department, University of Toledo, Toledo, OH 43606-3390}
\altaffiltext{2}{Space Telescope Science Institute, 3700 San Martin Drive, Baltimore, MD 21218}

\vspace{.2in}

\begin{abstract}
We study the stellar content in the tidal tails of three nearby
merging galaxies, NGC 520, NGC 2623, and NGC 3256, using \emph{BVI}
imaging taken with the Advanced Camera for
Surveys on board the \emph{Hubble Space Telescope}. The tidal tails in
all three systems contain compact and
fairly massive young star clusters, embedded in a sea of diffuse, unresolved stellar
light. We compare the measured colors and luminosities with predictions from population
synthesis models to estimate cluster ages and find that clusters began
forming in tidal tails during or shortly after
  the formation of the tails themselves.  We find a lack of very young
  clusters ($\le 10$ Myr old), implying that eventually star formation shuts off in the
  tails as the gas is used up or dispersed.  There are a few clusters
  in each tail with estimated ages that are older than the
modeled tails themselves, suggesting that these may have been stripped out
from the original galaxy disks.  The luminosity function of the tail clusters can be described by a single
power-law, $dN/dL \propto L^\alpha$, with $-2.6 < \alpha < -2.0$.  We find a stellar age gradient across some of the tidal tails, which we
interpret as a superposition of 1) newly formed stars and clusters along
the dense center of the tail and 2) a sea of broadly distributed, older stellar
material ejected from the progenitor galaxies.
\end{abstract}


\section{INTRODUCTION}

Tidal tails form during the interaction between galaxies, and are
composed of stars, gas, and dust ejected from the original galactic disks.  These interactions are often 
accompanied by one or more bursts of star and cluster formation in the main
bodies of the parent galaxies, according to simulations (e.g., Mihos,
Bothun, \& Richstone 1993) and
observations of star clusters (e.g., Whitmore \& Schweizer 1995).

The star cluster formation processes that occur within tidal tails,
however, are not well understood.  For example, building on pioneering
work by Knierman et al. (2003), Mullan et al. (2011; hereafter M11)
used V and $I$ band
photometry from Wide Field Planetary Camera 2 (WFPC2) on the \emph{Hubble Space Telescope}
(\emph{HST}) to measure the brightness and colors of point-like sources both in and
out of the tail region for 23 tidal tails.  They found a statistically significant number
of clusters in 10 out of their 23 tails, and they concluded that the
presence of tail clusters depends on several factors, including tail
age, \ion{H}{1} density, and surface brightness of the tail.  In this work we revisit
two of these tails using higher quality data taken with the
Advanced Camera for Surveys
(ACS)\footnote{http://www.stsci.edu/hst/acs} on \emph{HST}/Wide Field
Camera (WFC) and detect clusters in both of them for the first time.

Even less is known about the ages of star clusters in tidal tails.  de
Grijs et al. (2003) found $\sim 40$ clusters each in the Tadpole galaxy
with rough ages of $175 \pm 25$ Myr and in the Mice galaxies (NGC
4676) with ages $\approx 100 \pm 20$ Myr.  Most of these clusters
appear to have formed in the tails since their estimated ages are
younger than those of the tails;
de Grijs et al. estimated a dynamical age of
$400-800$ Myr for the Tadpole, and Privon et al. (2013) suggested a formation
time of $\sim 175$ Myr ago for the Mice.  Tran et al. also
observed the Tadpole galaxy, and found 42 extremely young tail clusters ($\sim
4-5$ Myr; 2003).  Bastian et al. (2005) found young massive star clusters in the
tails of NGC 6872 and was able to estimate ages and masses for
individual clusters using $UBVI$ and H$\alpha$ photometry.  No star
clusters older than the dynamical age of the tidal
tail were found (at least down to their completeness
limit of $M_V=-10.4$), once again suggesting that the
clusters formed from the tidal debris.  They also found a large population of
clusters that are very young ($< 10$ Myr).  While looking for spatial
trends, Bastian et al. found that older tail clusters tend to be
located closer to the main bodies, while younger clusters
are spread throughout the tail.  They argue that these findings suggest
an initial burst of cluster formation as the tails form.  After this burst,
as the tails expand, the gas clouds cool and condense,
eventually becoming sites of individual cluster formation.  

While tidal tails provide a unique and interesting environment to
study star clusters, they also provide key information for improving
simulations of interacting/merging galaxies.  Efforts to
reproduce mergers using simulations focus on tidal tail morphology
to constrain details of the galaxy interaction.
Ages of star clusters in tails provide constraints on the merger timescale.
The spatial distributions of clusters in the main bodies and tails
helps to discriminate between different prescriptions for star
formation within the simulations.  For
example, dynamical models of the interacting system NGC 7252 show that
density-dependent star formation results in a more
centrally concentrated distribution of main-body clusters, while
shock-induced star formation tends to spread cluster formation
throughout the galaxy, including in the tails (Chien \& Barnes 2010).

The goal of this work is to better understand how clusters form
and are distributed during the merging of two galaxies.  We focus our study on three
merging systems, NGC 520, NGC 2623, and NGC 3256.  
This paper is set up in the following manner.  In Section 2 we discuss the
observations and photometry, as well as cluster selection.  Section 3
presents the cluster densities and cluster luminosity functions, as well
as colors and ages of clusters and of the diffuse
light.  In Section 4 we discuss the galaxy merger history and
interpret our results. Section 5 summarizes the work and presents conclusions.

\section{OBSERVATIONS, DATA REDUCTION, AND CLUSTER SELECTION}

\subsection{Observations and Data Reduction}

\begin{figure*}[ht]
\centering$
\begin{array}{ccc}
\includegraphics[scale=0.50]{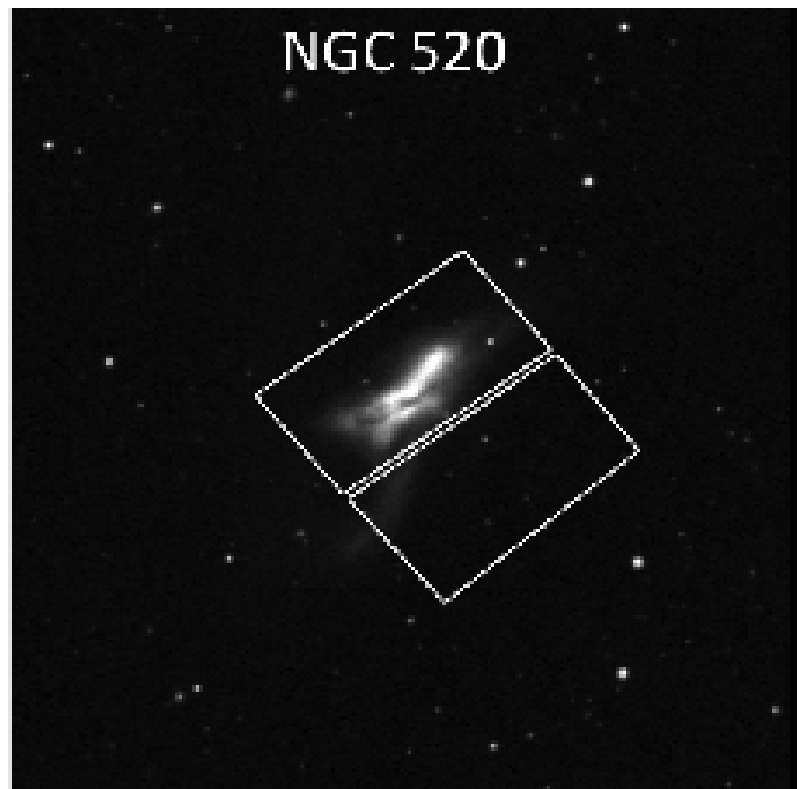}&
\includegraphics[scale=0.50]{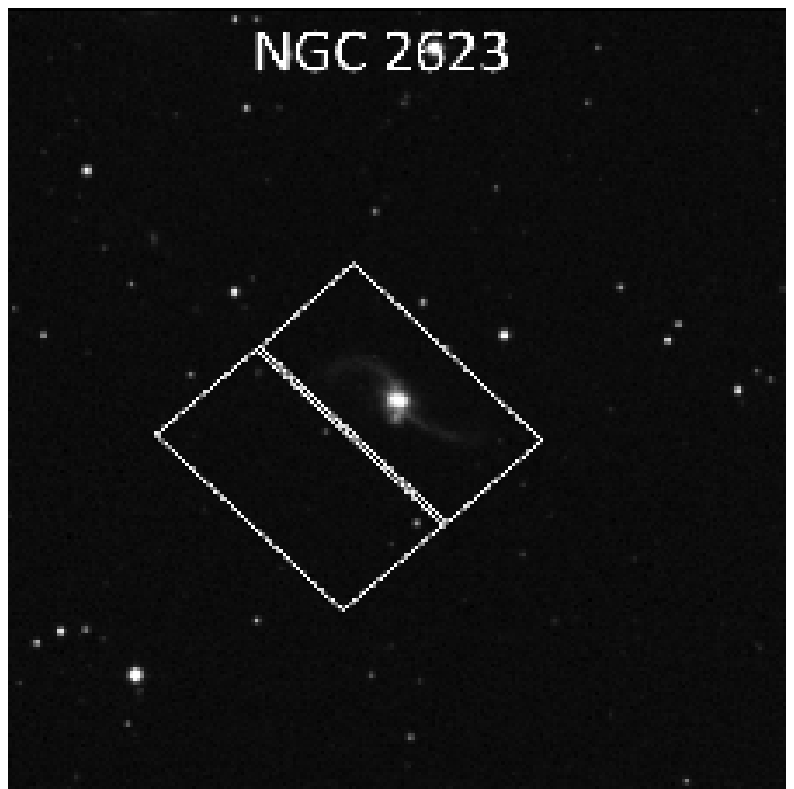}&
\includegraphics[scale=0.50]{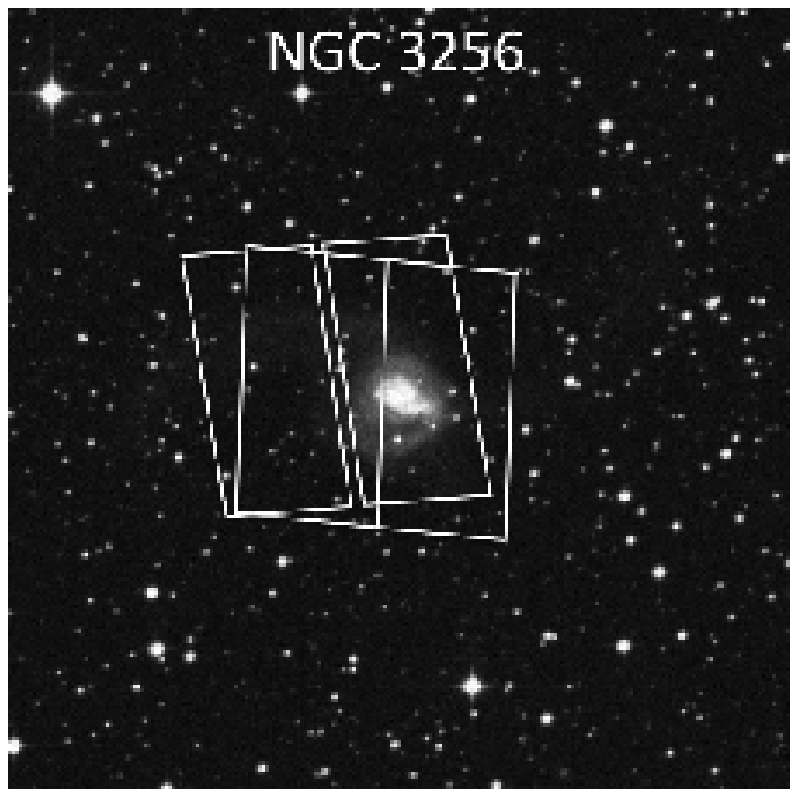}
\end{array}$
\caption{10' x 10' images of our galaxy merger sample from the
  Digitized Sky Survey.  Overlaid in each image is the WFC field of
  view.  In these and all the following images, north is up and east is to the left.}
\label{FOV}
\end{figure*}

\begin{figure*}[ht]
\centering$
\begin{array}{ccc}
\includegraphics[scale=0.50]{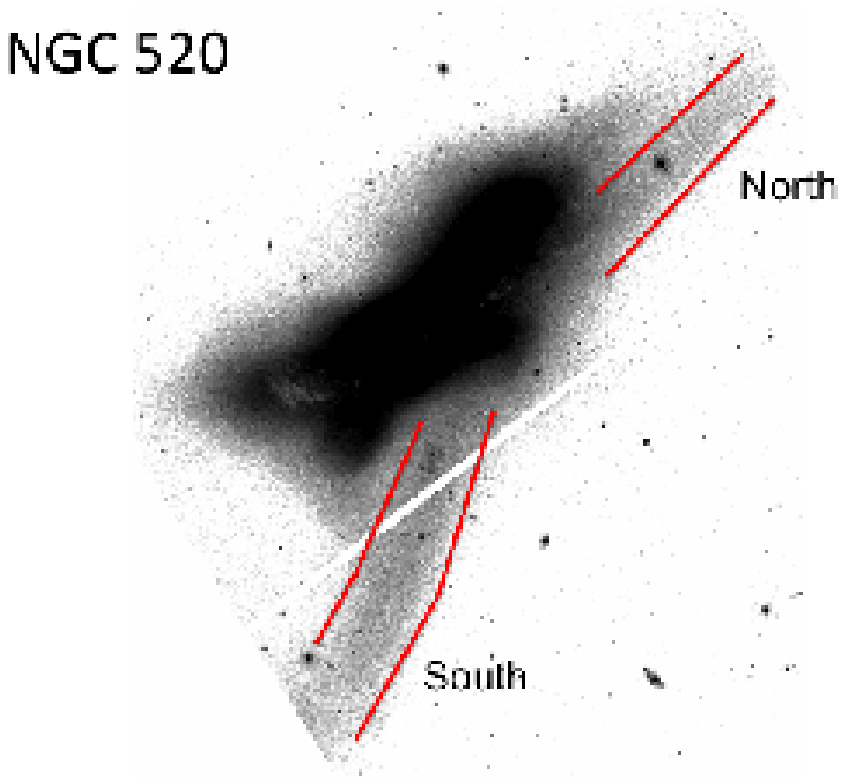}&
\includegraphics[scale=0.50]{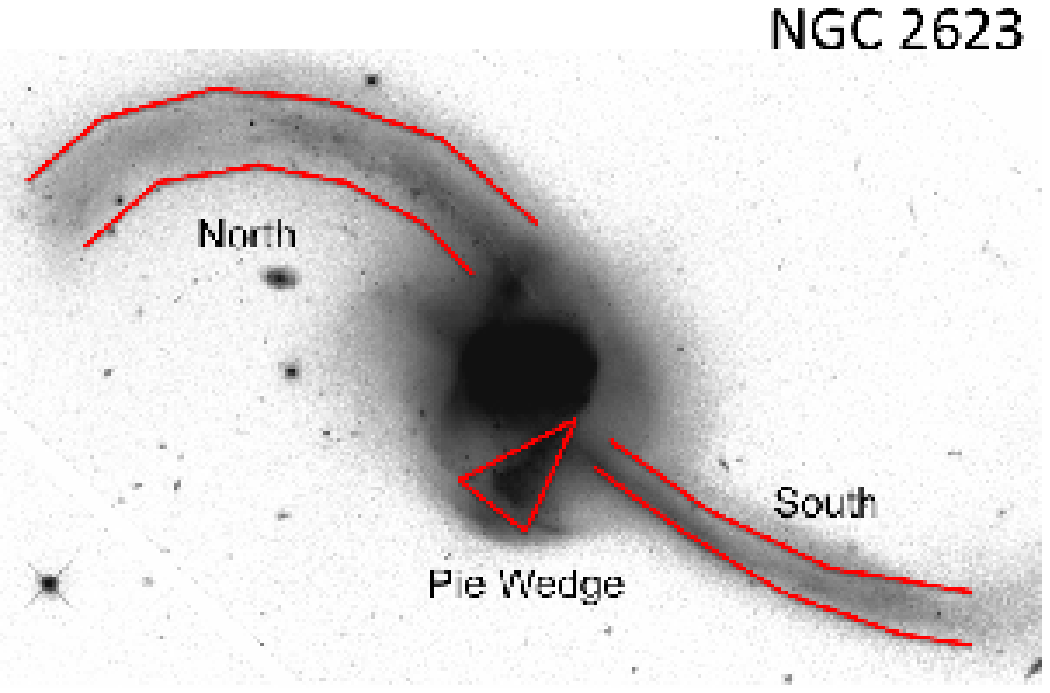}&
\includegraphics[scale=0.50]{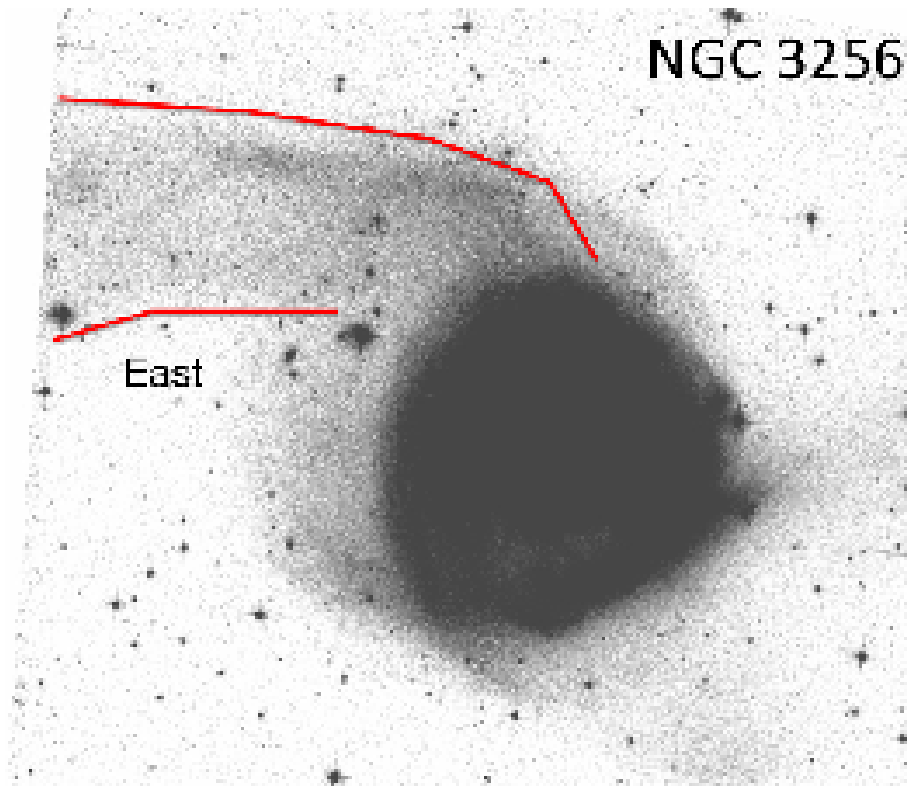}
\end{array}$
\caption{ACS/WFC $F814W$ images showing the sample of all three galaxies and their tails.}
\label{sample}
\end{figure*}

\begin{table*}
\caption{Galaxy Properties}
\label{Tab:galaxies}
\begin{center}
\begin{tabular}{l|cccccc}
\hline\hline								
Galaxy & R.A. & Dec. & Distance Modulus\footnote{Distance moduli come from
  distances obtained from recession velocity measurements relative to
  the local group, adopting $H_0 = 70$ km s$^{-1}$ Mpc$^{-1}$.} & $A_V$ & $E(B-V)$ & $E(V-I)$ \\
\hline
NGC 520  	& 01 24 35.1  &+03 47 33 & 32.46 & 0.077 & 0.025 & 0.035\\
NGC 2623	& 08 38 24.1 & +25 45 17 & 34.50 & 0.113 & 0.036 & 0.051\\
NGC 3256	& 10 27 51.3 & -43 54 13 & 32.79 & 0.334 & 0.107 & 0.151\\
\hline
\end{tabular}
\end{center}
\end{table*}

Observations of NGC 520 and NGC 2623 were taken with the WFC on
the ACS on \emph{HST} as part of program
GO-9735 (PI: Whitmore). NGC 520 was observed in 2004 October using the filters \emph{F435W}
($\approx B$ in the Johnson-Cousins system; exposed for 2400 s),
\emph{F555W} ($\approx V$; 1000 s),
and \emph{F814W} ($\approx I$; 1200 s).  The entire galaxy, including
the majority of its tidal tails, were covered in a single WFC frame.  Observations of NGC
2623 were taken in 2004 February using the same filters with exposure
times of 3730, 1206 and 2460 s, respectively.  It was also covered
in a single WFC frame.  Images of NGC 3256 come from two
separate observing proposals.  \emph{F555W} filter observations of NGC 3256 were
obtained in 2003 November as part of the program GO-9735 (2552 s).
WFC observations using \emph{F435W} and \emph{F814W} filters were
taken in 2005 November as part of program GO-10592 (PI: Evans) for
1320 and 760 s, respectively.  Both fields cover most of NGC 3256's eastern
tail, as well as its main body.  Figure \ref{FOV} shows the WFC fields
of view used in this study.

The raw data were processed through the standard ACS pipeline.
The reduced, multidrizzled images were taken from the Hubble Legacy
Archive and have a scale of 0.05$''$ per pixel.  Figure \ref{sample}
shows \emph{F814W} images of our targets, with the tidal
tail portions studied in this work labeled.  We will refer to specific
tails within a galaxy by abbreviating the location of each tail
(i.e. the northern tail of NGC 520 will be called NGC 520N).  We
define the tail edges as the locations where the counts drop to
3$\sigma$ above the general background level.

Figure \ref{sample} shows that NGC 520 has two tidal tails, commonly referred to
as the north and south tails.  Both of these
tails extend beyond the field of view in Figure \ref{sample}.  Not seen in Figure
\ref{sample} is the dwarf galaxy UGC 957, at about $6'$ north-west
of NGC 520.  NGC 2623 is in the middle panel of
Figure \ref{sample}.  It exhibits two nearly symmetric tidal tails,
although the northern tail is wider than the
southern tail.  An extremely bright, triangle-shaped region south of the nucleus of the
galaxy, which we call the pie wedge, is also highlighted.  NGC 3256 is
on the right in Figure \ref{sample}.  While it
has two tails (east and west), only part of the eastern tail was
observed.  The three mergers are next to each other on the Toomre
Sequence, in the order NGC 520, NGC 2623, and NGC 3256 (Toomre 1977).
General properties of these systems are listed in Table
\ref{Tab:galaxies}.  

\subsection{Cluster Selection}

Star cluster candidates were found by running the IRAF\footnote{IRAF
  is distributed by the National Optical Astronomy Observatory, which
  is operated by the Association of Universities for Research in
  Astronomy (AURA) under a cooperative agreement with the National Science Foundation.} task
DAOFIND on the $I$ band images.  A few thousand objects were detected
in each galaxy.  We perform
$BVI$ aperture photometry on all sources in NGC 520 and NGC 3256
using an aperture size of 3 pixels in radius and a background area of
radii 8 to 11 pixels using the IRAF task PHOT.
Because of the greater distance to NGC 2623, we use a smaller
  aperture of 2 pixels (with the same background area) in order to minimize
contamination from nearby sources.  In the pie wedge region, we
estimate the background level using annuli between
5--8 pixels, in order to minimize the impact of crowding.  Magnitudes
were converted to the
VEGAMAG system using zeropoints calculated from the ACS zeropoint
calculator\footnote{http://www.stsci.edu/hst/acs/analysis/zeropoints}.
Aperture corrections were performed in NGC 3256 by selecting the
brightest $\sim 20$ isolated clusters in each filter and finding their mean
3--10 pixel magnitude difference.  We then applied the 10 pixel to infinity
aperture corrections taken from the encircled energy catalog for
ACS/WFC in Table 3 of Sirianni et al. (2005).  The final corrections
were 0.45, 0.42, and 0.54 magnitudes for \emph{B}, \emph{V}, and
\emph{I}, respectively, and they were applied to each cluster.
Unfortunately, the background level was too high in NGC 520 and
NGC 2623 for reliable aperture correction determinations.  We find that clusters in NGC
520 have a similar range of sizes as those in NGC 3256 and therefore applied the
aperture corrections derived from NGC 3256 to the NGC 520 clusters.
Clusters in the more distant NGC 2623 are nearly indistinguishable from point
sources; we assume the 2 pixel to infinity aperture corrections from Table 3
of Sirianni et al. (2005).

We measure FWHM values for all detected sources using the
\emph{ISHAPE} software (Larsen 1999).  \emph{ISHAPE}
measures FWHMs by convolving the point spread function with a
King profile, then performing a $\chi^2$ calculation to test the
goodness of fit to each individual cluster (King 1966).  \emph{ISHAPE}
iterates through different values for the effective radius until a
minimum $\chi^2$ is found.  

We select cluster candidates using a combination of automated
selection criteria (listed in Table \ref{Tab:cuts}) and visual
inspection of each candidate (see Bastian et al. 2012 for further discussion on the
``hybrid'' method).  The selection criteria in $m_{V}$, FWHM,
and magnitude error ($\sigma_V$) remove most foreground stars and
background galaxies.  We made a cut at the bright end of $m_V=20$, in
order to eliminate saturated foreground stars (brighter than any
cluster expected in these mergers).  The faint
end of our cluster catalog likely suffers from incompleteness.  We
restrict our sample to luminosities brighter than the point where
the luminosity function (discussed in Section 3.2) turns down rather
than continuing to increase in a power-law fashion.  We use cuts in
FWHM (size) to separate stars and background galaxies from clusters,
to the highest degree possible.  In NGC 3256, we exclude
sources with FWHM values less than 0.2 pixels, because the FWHM of the
vast majority of likely field stars was measured to be smaller than
that value. We also place an upper limit of 2.0
pixels in FWHM to exclude very extended background galaxies.  The
lower limit on FWHM for
clusters in NGC 520 was chosen to be 0.1 pixels because there is a
group of young, compact clusters in the northern portion of NGC
520S.  At the distance of NGC 2623, clusters are essentially point
sources, so FWHM measurements cannot distinguish clusters from
bright, individual stars.  Therefore, in addition to an upper limit on
FWHM of 2.0, we apply a cutoff in
magnitude uncertainty, $\sigma_V$.  In the tails, $\sigma_V <
0.1$ did a good job removing obvious non-clusters, but in the pie wedge,
we relax our $\sigma_V$ cut to 0.2 because the area is more
crowded and has a very bright background.

\begin{table}
\caption{Selection Cuts}
\label{Tab:cuts}
\begin{center}
\begin{tabular}{l|ccc}
\hline\hline								
Galaxy & $m_{V}$ & FWHM & $\sigma_V$\\
\hline
NGC 520  	& 25.0 & 0.1 - 2.0 & \nodata \\
NGC 2623	& 26.5 & $\leq 2.0$ & 0.1 \\
pie wedge	& 26.5 & $\leq 2.0$ & 0.2 \\
NGC 3256	& 26.0 & 0.2 - 2.0 & \nodata\\
\hline
\end{tabular}
\end{center}
\end{table}

\begin{figure}[h]
\centering$
\begin{array}{cccc}
\includegraphics[scale=0.30]{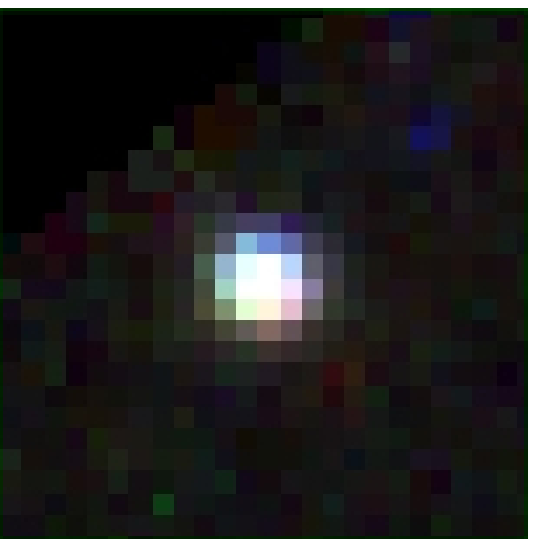}&
\includegraphics[scale=0.30]{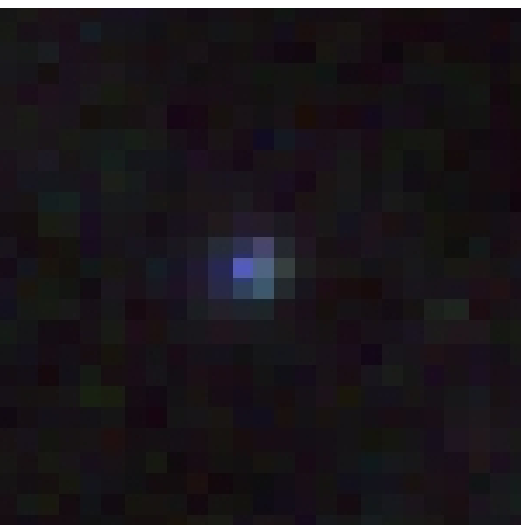}&
\includegraphics[scale=0.30]{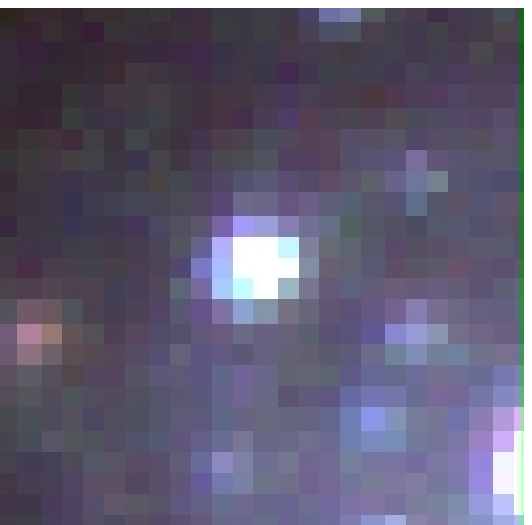}&
\includegraphics[scale=0.30]{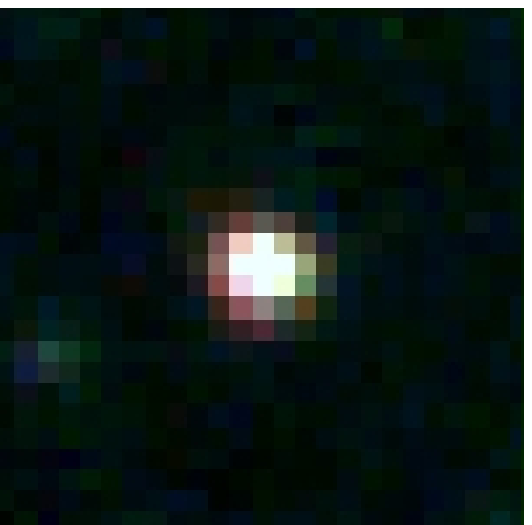}\\
\includegraphics[scale=0.30]{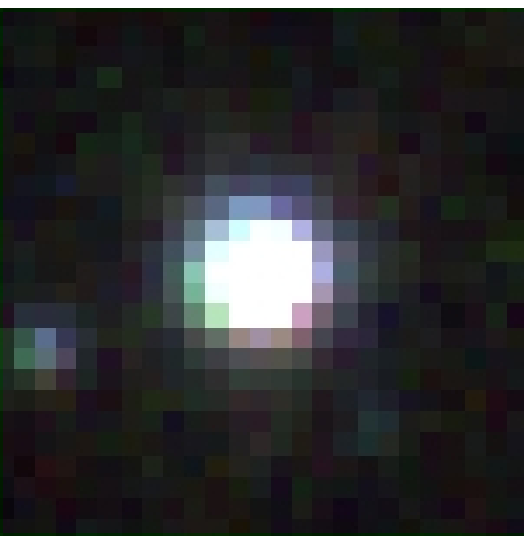}&
\includegraphics[scale=0.30]{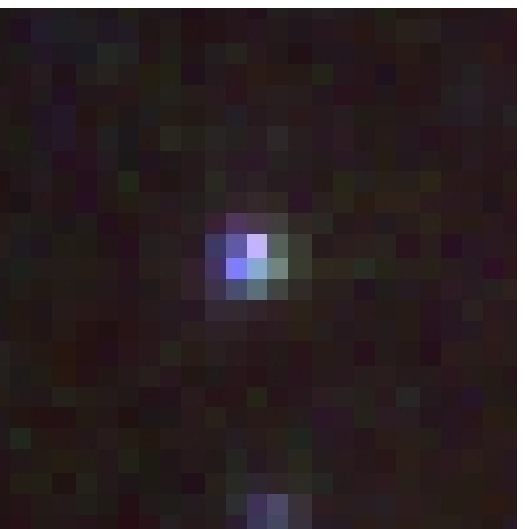}&
\includegraphics[scale=0.30]{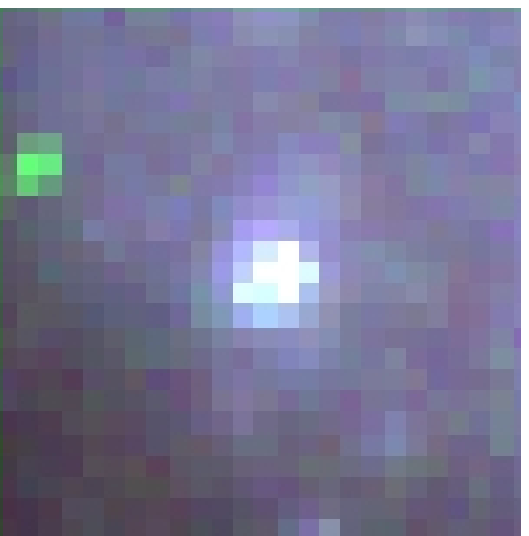}&
\includegraphics[scale=0.30]{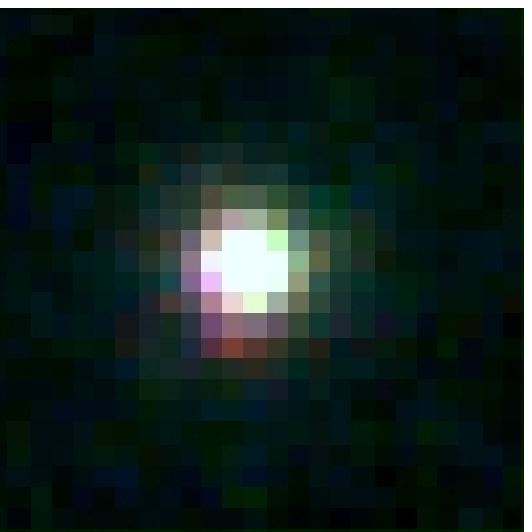}\\
\includegraphics[scale=0.30]{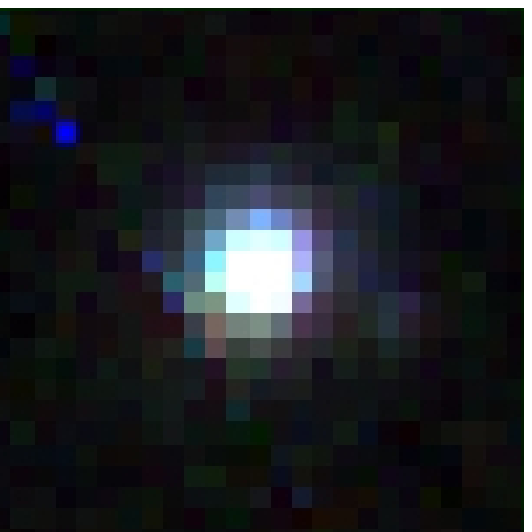}&
\includegraphics[scale=0.30]{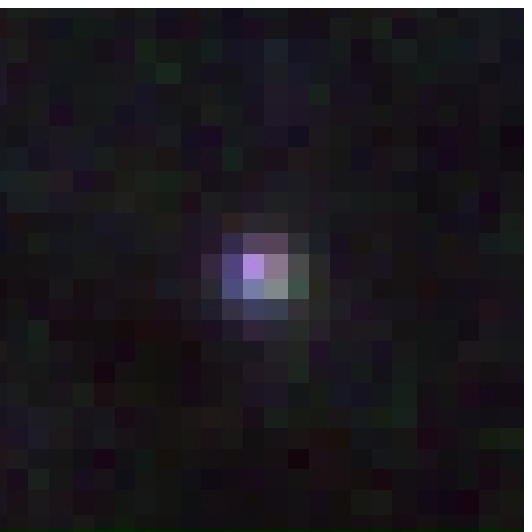}&
\includegraphics[scale=0.30]{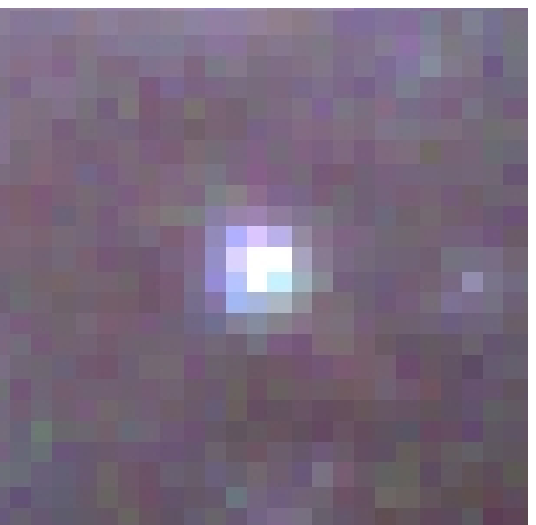}&
\includegraphics[scale=0.30]{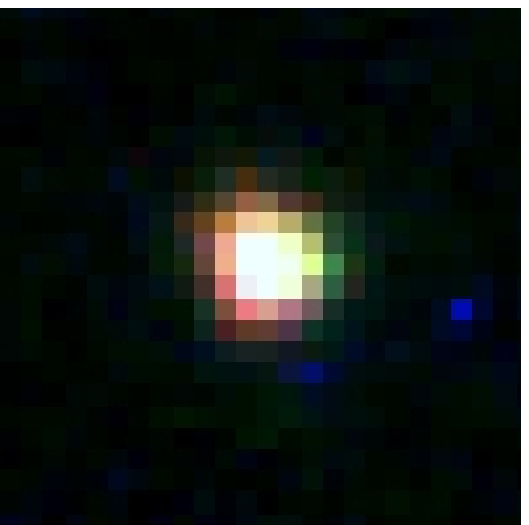}\\
\end{array}$
\caption{Some of the brightest star clusters in the tidal tails of each galaxy.  Each
  image is 25 x 25 pixels.  Galaxies are separated by column from left to
  right, in the order NGC 520, NGC 2623 (tail region), NGC 2623 (pie
  wedge region), and NGC 3256.}
\label{thumbs}
\end{figure}

\begin{figure*}[ht]
\centering$
\begin{array}{c}
\includegraphics[scale=0.65]{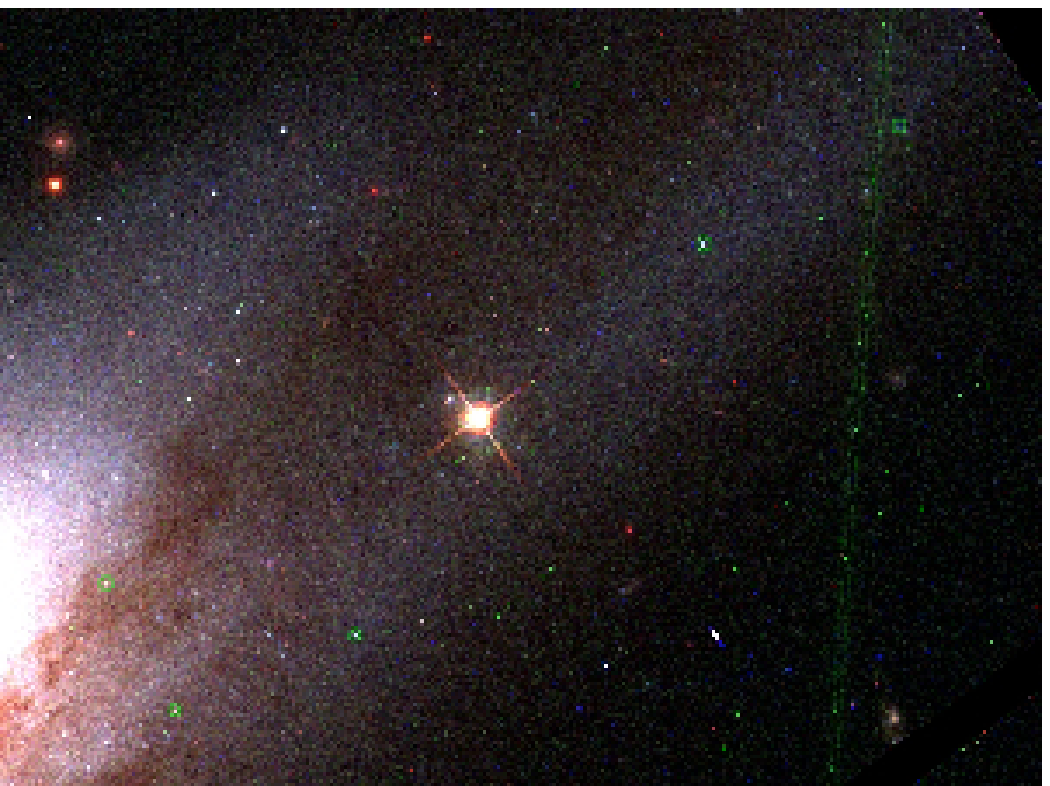} \\
\end{array}$
$\begin{array}{cc}
\includegraphics[scale=0.50]{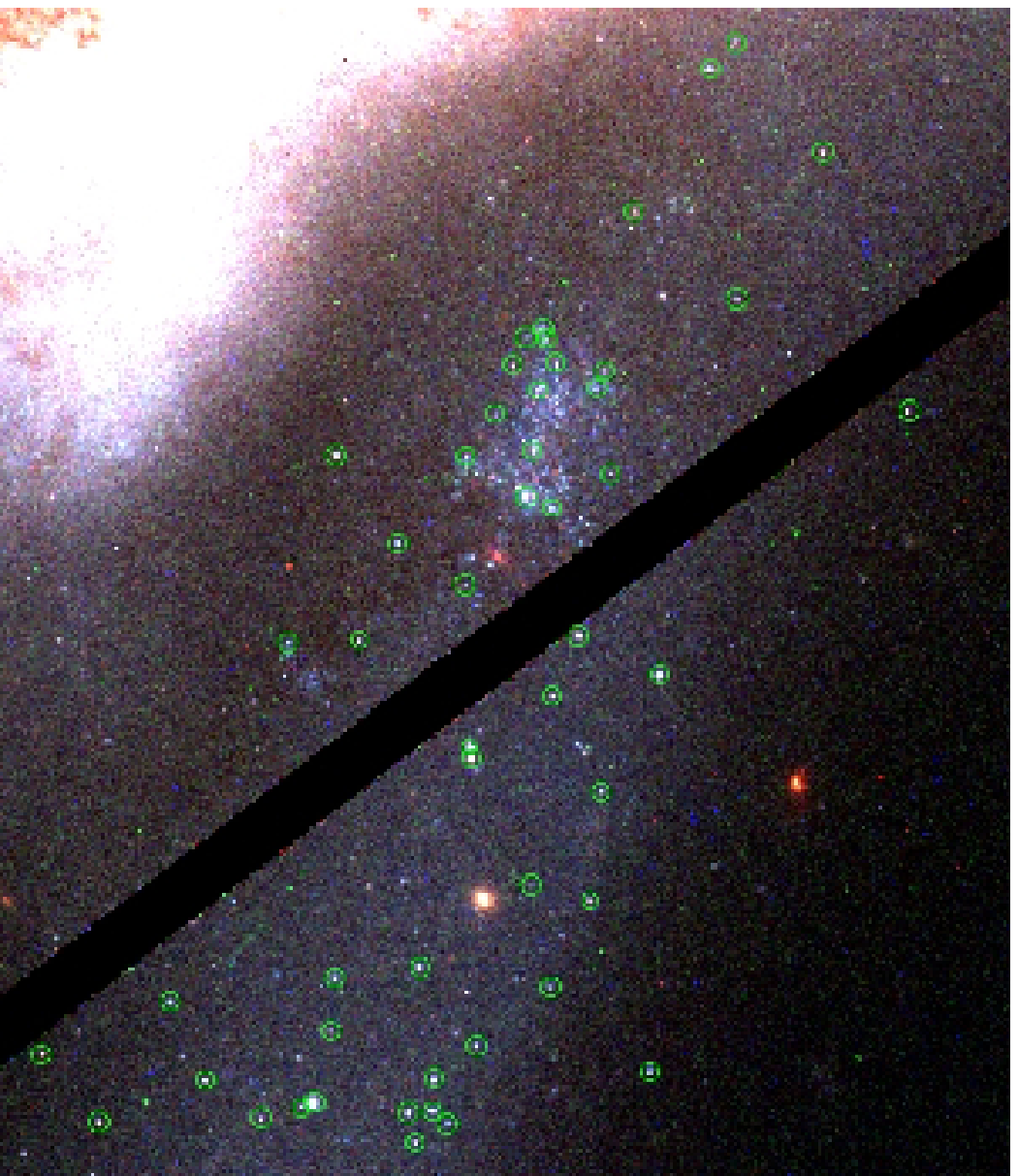}& 
\includegraphics[scale=0.50]{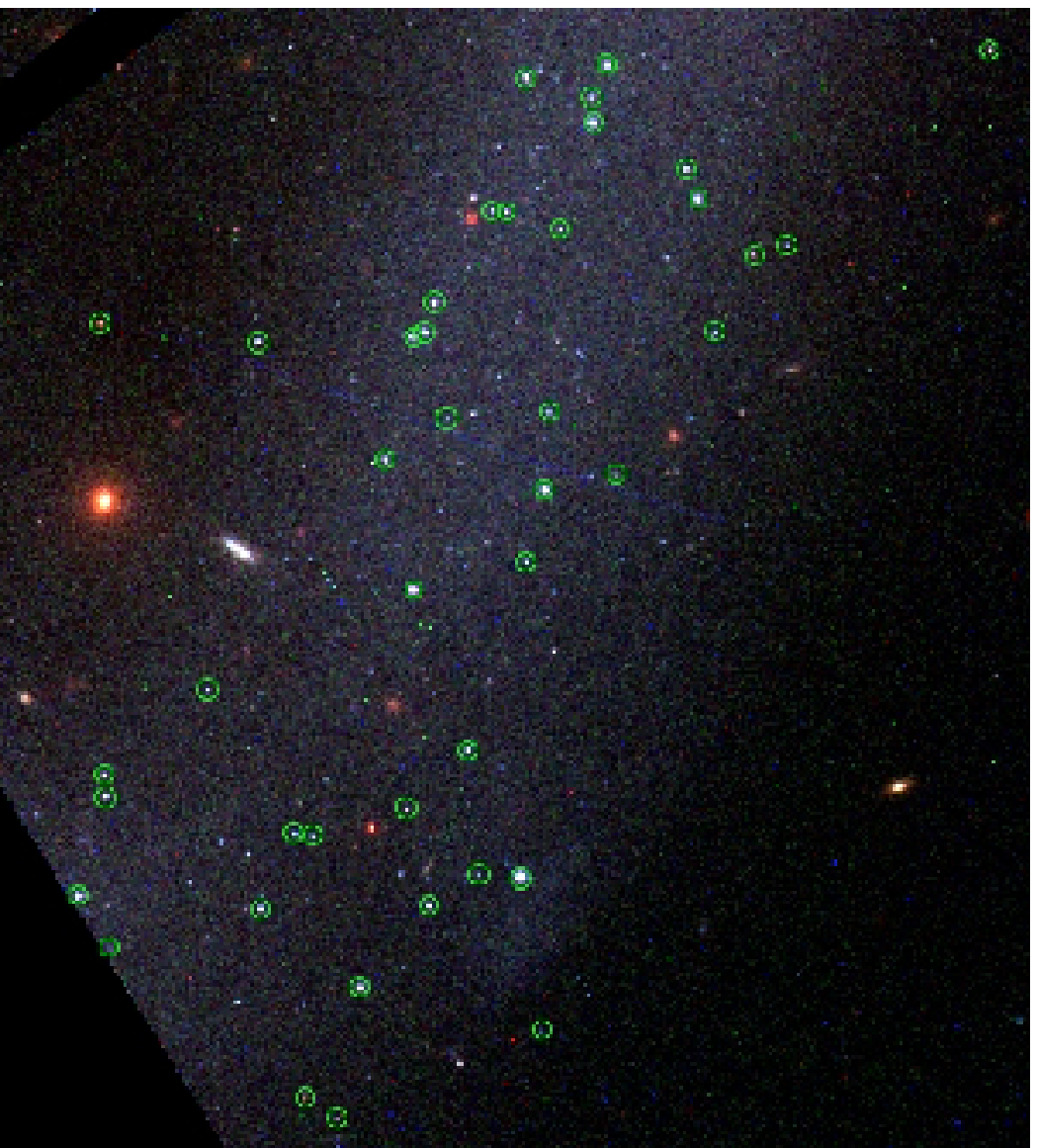}
\end{array}$
\caption{NGC 520 cluster candidates are marked with green circles.  On the top is the northern tail.  The
  lower-left panel shows the northern portion of the southern tail.  On
  the lower-right is the southern half of the southern tail.}
\label{520_tailreg}
\end{figure*}

\begin{figure*}[h]
\centering$
\begin{array}{c}
\includegraphics[scale=0.55]{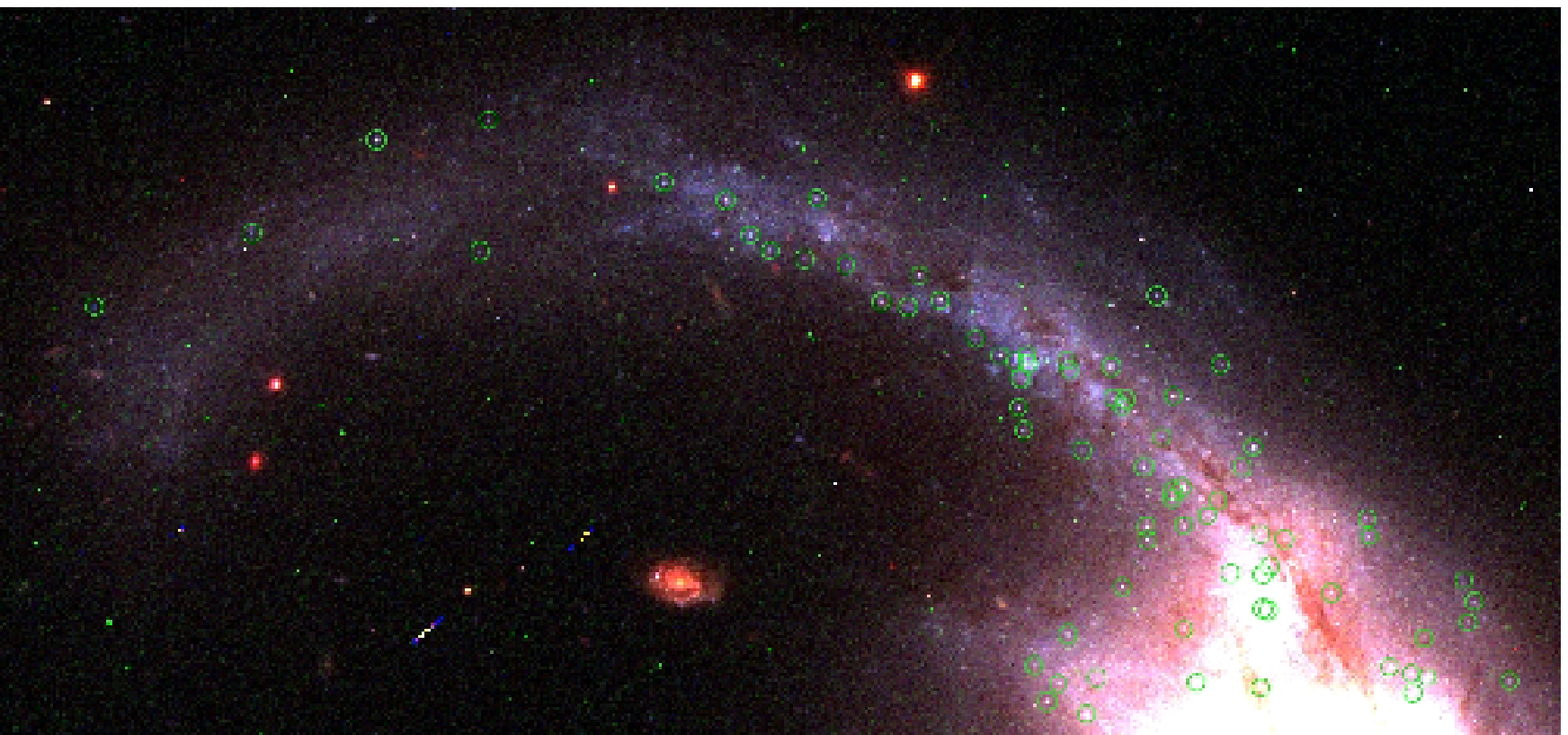}\\
\includegraphics[scale=0.55]{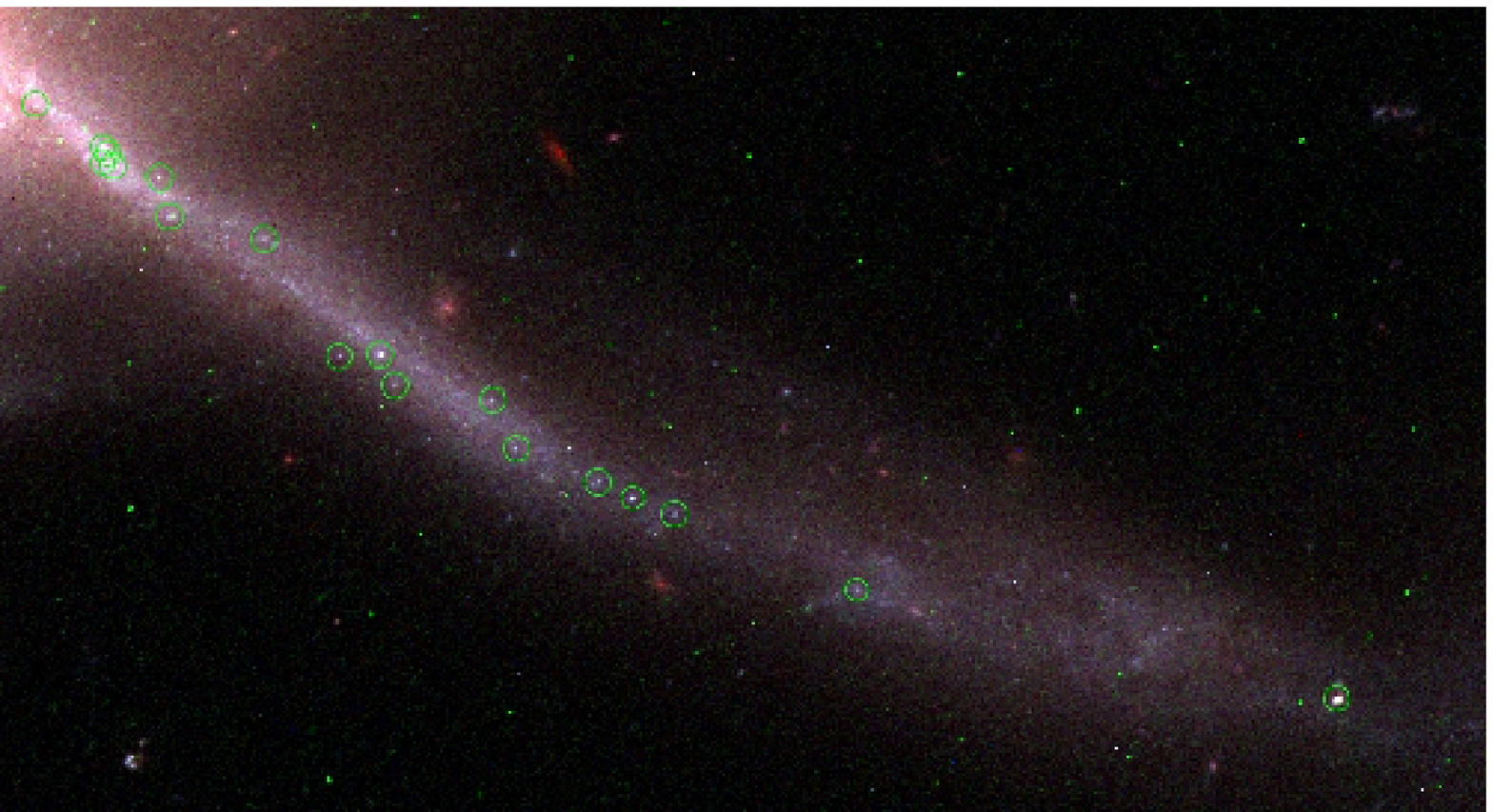}
\end{array}$
\caption{NGC 2623 cluster candidates.  On top is the
  northern tail.  The southern tail is shown on bottom.}
\label{2623_tailreg}
\end{figure*}

\begin{figure*}[h]
\centering
\includegraphics[scale=0.90]{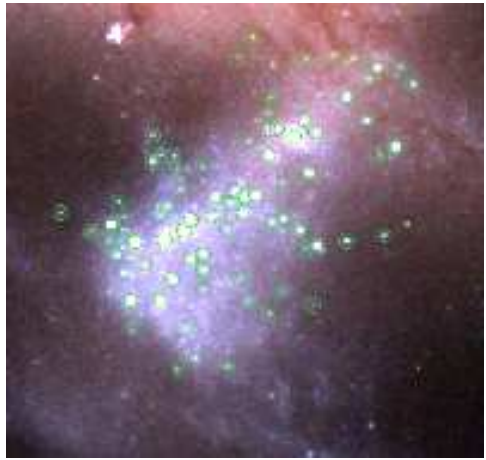}
\caption{NGC 2623 pie wedge cluster candidates.}
\label{2623_piereg}
\end{figure*}

\begin{figure*}[hb]
\centering
\includegraphics[scale=0.60]{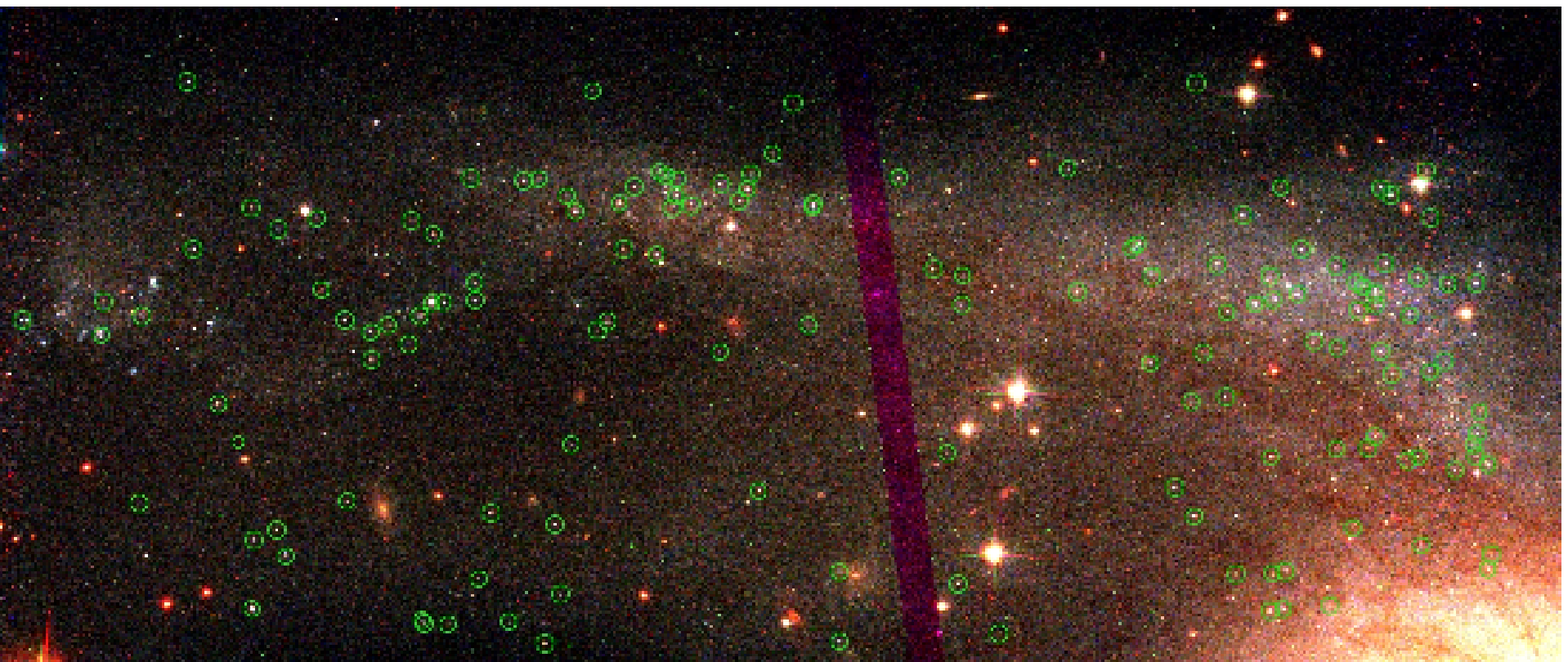}
\caption{NGC 3256E cluster candidates.}
\label{3256_tailreg}
\end{figure*}

Finally, we visually inspect each cluster candidate. For each galaxy
we identify high signal-to-noise ration (S/N) clusters, as
well as obvious foreground stars, as benchmarks for fainter objects.  We choose clusters based on
their fuzzy appearance, as well as wider radial profiles than those of
stars.  Figure \ref{thumbs} shows three of
the brightest clusters from each galaxy's tidal tails, and Figures
\ref{520_tailreg} -- \ref{3256_tailreg} show locations of clusters in
each tail.  All possible clusters were identified independently of the
selection cuts listed above, then later compared to the sample
obtained from the selection cuts.  We find excellent agreement between
the two methods in NGC 520 and NGC
3256, and we only include sources that both fit our selection cuts and
visually appear as clusters.  Due to the poor S/N in NGC 2623, several
sources appeared visually as likely clusters but had FWHM and/or
$\sigma_V$ values outside of the selection criteria.  Because of this
low S/N, we rely more heavily on visual inspection than selection
cuts and included some sources with FWHM or $\sigma_V$ values
outside of the cuts.  The total number of clusters found in each tidal
tail ($N_{clus}$) is compiled in Table \ref{Tab:clusters}, and
  we compile the cluster catalog in Table \ref{Tab:catalog}.

\begin{table*}
\footnotesize
\caption{Properties of star clusters.  $\tau_{\rm{tail}}$ [Myr] is the predicted
  age of the tidal tail as estimated from simulations.  $N_{\rm{clus}}$ is the total number of
  clusters found in the region.  $\Sigma_{\rm{clus}}$ is the star cluster
  density per kpc$^2$ where $M_{V} < -8.5$.  $\tau_{\rm{clus}}$ [Myr] is
  the approximate age of the bulk cluster formation.
  Median colors ($\langle$B - V$\rangle$,
  $\langle$V - I$\rangle$) are for clusters \emph{younger} than the
  age of the tidal tail.  $\alpha$ values (from the relation $dN/dL
  \propto L^{\alpha}$) listed come from the
  $\alpha_{\rm{const-num}}$ method.}
\label{Tab:clusters}
\begin{center}
\begin{tabular}{lcc|cccccc}
\hline\hline						
Galaxy & Tail & $\tau_{\rm{tail}}$ & $N_{\rm{clus}}$ &
$\Sigma_{\rm{clus}}$ & $\tau_{\rm{clus}}$ & $\langle$B - V$\rangle$ &
$\langle$V - I$\rangle$ & $\alpha$\\
\hline
NGC 520  	& North  & 300 & 5 & 0.00 & $\sim 100$ & 0.08 $\pm$ 0.05 &
0.38 $\pm$ 0.05 & \nodata \\
NGC 520	        & South  & 300 & 93 & 0.49 & 260 & 0.19
$\pm$ 0.09 & 0.60 $\pm$ 0.17 & -2.32 $\pm$ 0.36\\
NGC 2623	& North  & 220 & 75 &0.54 & 230 & 0.10
$\pm$ 0.15 & 0.48 $\pm$ 0.28 & -2.39 $\pm$ 0.34\\
NGC 2623	& South  & 220 & 18 & 0.32 & \nodata & 0.12
$\pm$ 0.13 & 0.43 $\pm$ 0.17 & \nodata\\
NGC 2623	& pie wedge  & $\le 220$ & 76 & 8.5 & $\sim 100$ &
0.05 $\pm$ 0.15 & 0.43 $\pm$ 0.19 & -2.02 $\pm$ 0.21\\
NGC 3256	& East & 450 & 141 & 0.10 & $ \geq 260$ & 0.21
$\pm$ 0.12 & 0.58 $\pm$ 0.21 & -2.61 $\pm$ 0.27\\
\hline
\end{tabular}
\end{center}
\end{table*}

\begin{table*}
\caption{Cluster catalog.  Apparent magnitudes ($m_{\lambda}$) and photometric errors
  ($\sigma_{\lambda}$) are listed, as well as individual cluster ages
  in log years ($\tau$). Table \ref{Tab:catalog} is published in its entirety in the electronic
  edition of ApJ.  A portion is shown here for guidance regarding its form and content.}  
\label{Tab:catalog}
\begin{center}
\begin{tabular}{lcccccccccc}
\hline\hline								
Tail & ID & RA & Dec & $m_B$ & $\sigma_B$ & $m_V$ & $\sigma_V$ & $m_I$ & $\sigma_I$ & $\tau$\\
\hline
520N & 1 & 1 24 31.77 & +3 47 59.03 & 24.68 & 0.02 &  24.01 & 0.03 & 22.82 & 0.02 & 9.11 \\
520N & 2 & 1 24 31.02 & +3 48 3.70 & 24.63 &  0.03 & 24.47 &  0.04 & 24.10 &  0.04 & 8.06 \\
520N & 3 & 1 24 32.06 & +3 48 6.88 & 25.82 &  0.07 & 24.70 &  0.05 & 23.41 &  0.03 & 9.99 \\
520N & 4 & 1 24 29.58 & +3 48 27.96 & 24.31 &  0.02 & 24.23 &  0.03 & 23.85 &  0.03 & 8.06 \\
520N & 5 & 1 24 28.77 & +3 48 35.25 & 24.41 &  0.02 & 24.33 &  0.03 & 23.88 &  0.03 & 8.06 \\
\hline
520S & 1 & 1 24 36.18 & +3 45 37.42 & 24.95 &  0.03 & 24.56 &  0.04 & 23.76 &  0.03 & 8.76 \\
520S & 2 & 1 24 36.27 & +3 45 38.32 & 25.14 & 0.03 & 24.40 & 0.04 & 23.20 &  0.02 & 9.30 \\
520S & 3 & 1 24 35.59 & +3 45 41.18 & 25.00 & 0.03 & 24.94 & 0.05 & 24.34 & 0.04 & 7.59 \\
520S & 4 & 1 24 36.11 & +3 45 43.04 & 23.66 & 0.01 & 23.25 & 0.02 & 22.46 & 0.01 & 8.81 \\
\hline
\end{tabular}
\end{center}
\end{table*}

Distant, red spheroidal galaxies can have a similar color and appearance to
compact star clusters.  To assess the possible contamination from
background galaxies, we examine the colors and locations of sources
that made it through our automated cuts, but are located outside of
the tidal tail regions, with the reasonable expectation that most of these are
likely background galaxies.  However, we find no such objects in any
of our fields, only extended, obvious background galaxies.  In
addition, many background galaxies should appear elliptical rather
than circular.  Using \emph{ISHAPE}, we measure the ellipticity of
each cluster candidate and find few elongated red sources.  We
conclude that the fraction of background galaxies that have made it
through our cluster selection criteria is negligible.

\section{RESULTS}
We find a population of star clusters in each tidal tail in our sample
of galaxy mergers.  In this section we present the spatial densities,
luminosities, and colors of clusters in each tidal tail and estimate
their ages.

\subsection{Cluster Density}

One quantitative measure of a cluster population is its surface number
density, which is often used when cluster selection must be done
statistically.  We measure the number density of clusters,
$\Sigma_{\rm{clus}}$ (clusters kpc$^{-2}$),
for each tidal tail based on the catalogs in this work.  We choose a
magnitude cutoff of $M_V < -8.5$ (see Section 3.2 for details
  on absolute magnitude calculation) 
to directly compare our results to those of M11
(although altering the cutoff to $M_V = -8.0$ and $-9.0$ yields
qualitatively similar results).  NGC 520S yielded
$\Sigma_{\rm{clus}}=0.49$, while NGC 520N
yielded $\Sigma_{\rm{clus}} =0.0$ because its five clusters were all
fainter than $M_V=-8.5$.  The pie wedge in NGC 2623 possesses the highest
cluster density in our sample by over an order of magnitude at
$\Sigma_{\rm{clus}}=8.5$, while the north and south tails yielded
$\Sigma_{\rm{clus}} = 0.54$ and 0.32, respectively.  Lastly, NGC 3256E
possessed a relatively low density of $\Sigma_{\rm{clus}}=0.10$.  We find
that while all tidal tails in our sample contain clusters, there is a
large variation in $\Sigma_{\rm{clus}}$ for each tail.

For both NGC 520 and NGC 3256, we are able to directly compare
our values of $\Sigma_{\rm{clus}}$ to those values found in M11, although
we note that we cover different areas of the tails for both systems.
We also note that we adopt slightly
different distance moduli than M11, resulting in small differences in
$\Sigma_{\rm{clus}}$.  M11 found
$\Sigma_{\rm{clus}} = -0.011 \pm 0.013$ in NGC 520N and $\Sigma_{\rm{clus}} =
0.004 \pm 0.082$ for NGC 3256E, both of which are consistent with
zero.  While our results for NGC 520N agree with M11, we find a clear
population of clusters in NGC 3256E brighter than $M_V < -8.5$.  M11
also measured $\Sigma_{\rm{clus}}$ for $M_V < -7.5$ and $M_V < -6.5$,
however we do not compare $\Sigma_{\rm{clus}}$ values at these cutoffs due
to incompleteness.  We find that, in general, younger tails tend to
exhibit higher $\Sigma_{\rm{clus}}$, in agreement with the results of M11.

\subsection{Luminosity Functions (LFs)}

\begin{figure*}[ht]
\centering$
\begin{array}{cc}
\includegraphics[scale=0.45]{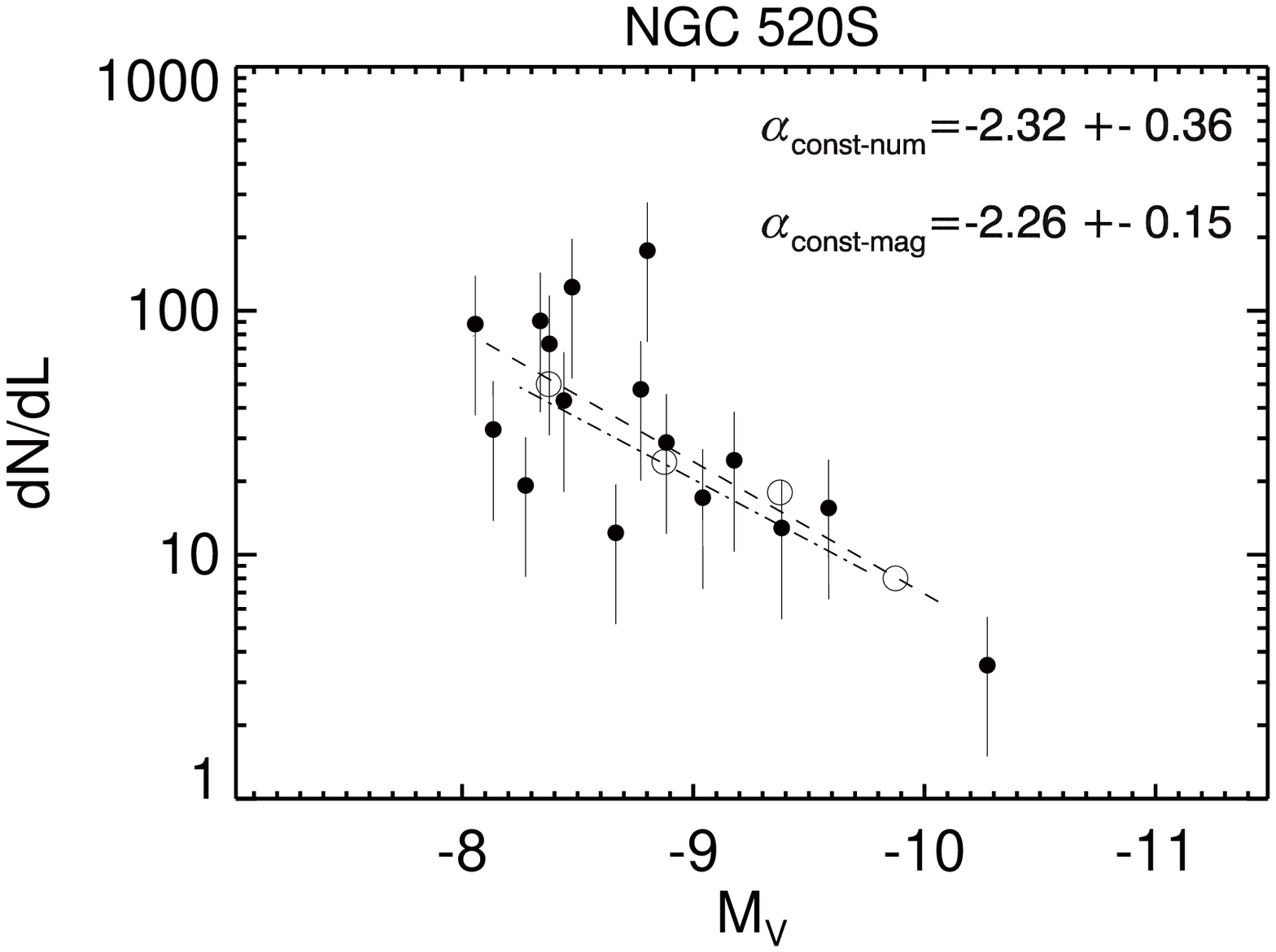}&
\includegraphics[scale=0.45]{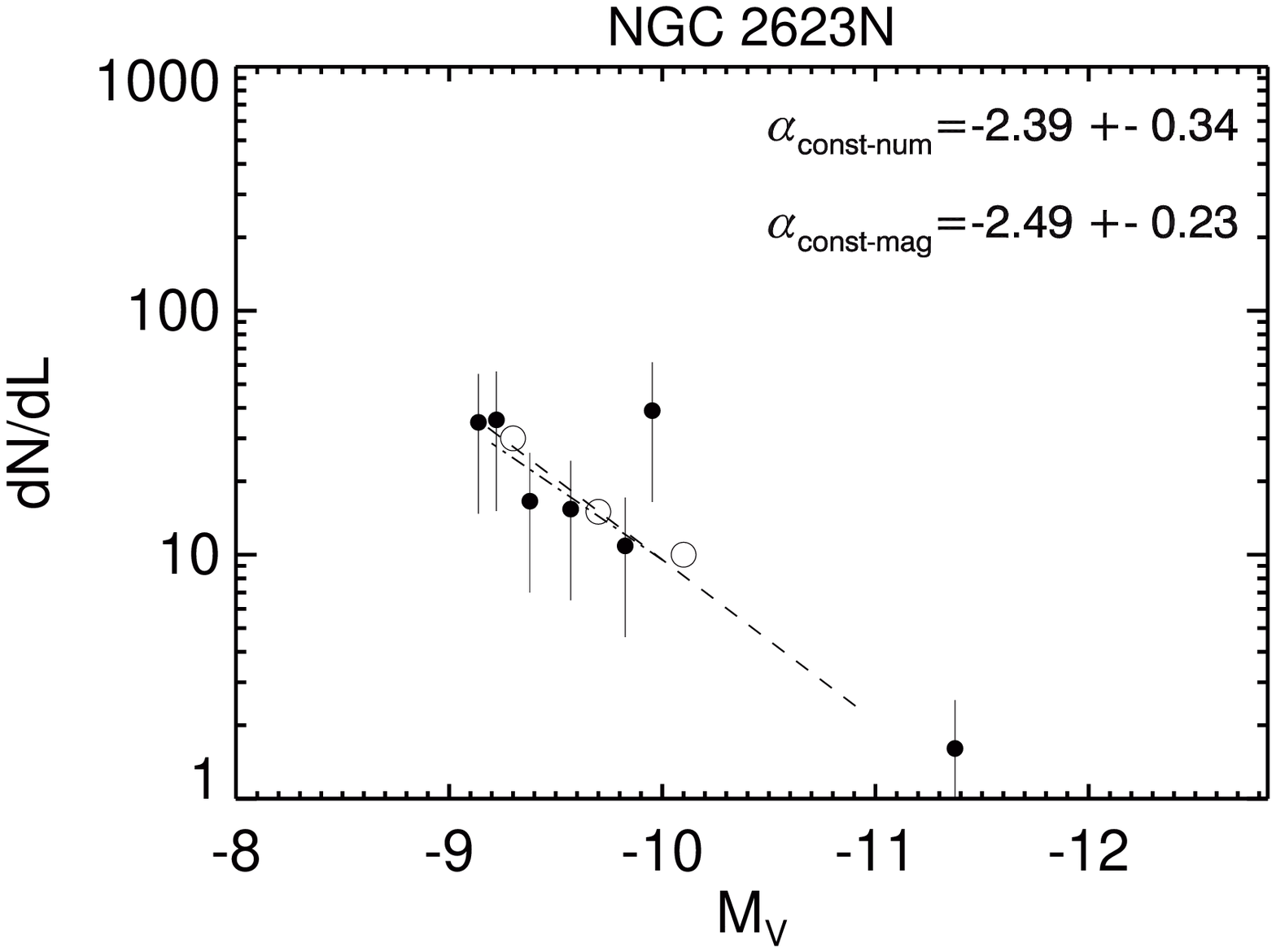}
\end{array}$
$\begin{array}{cc}
\includegraphics[scale=0.45]{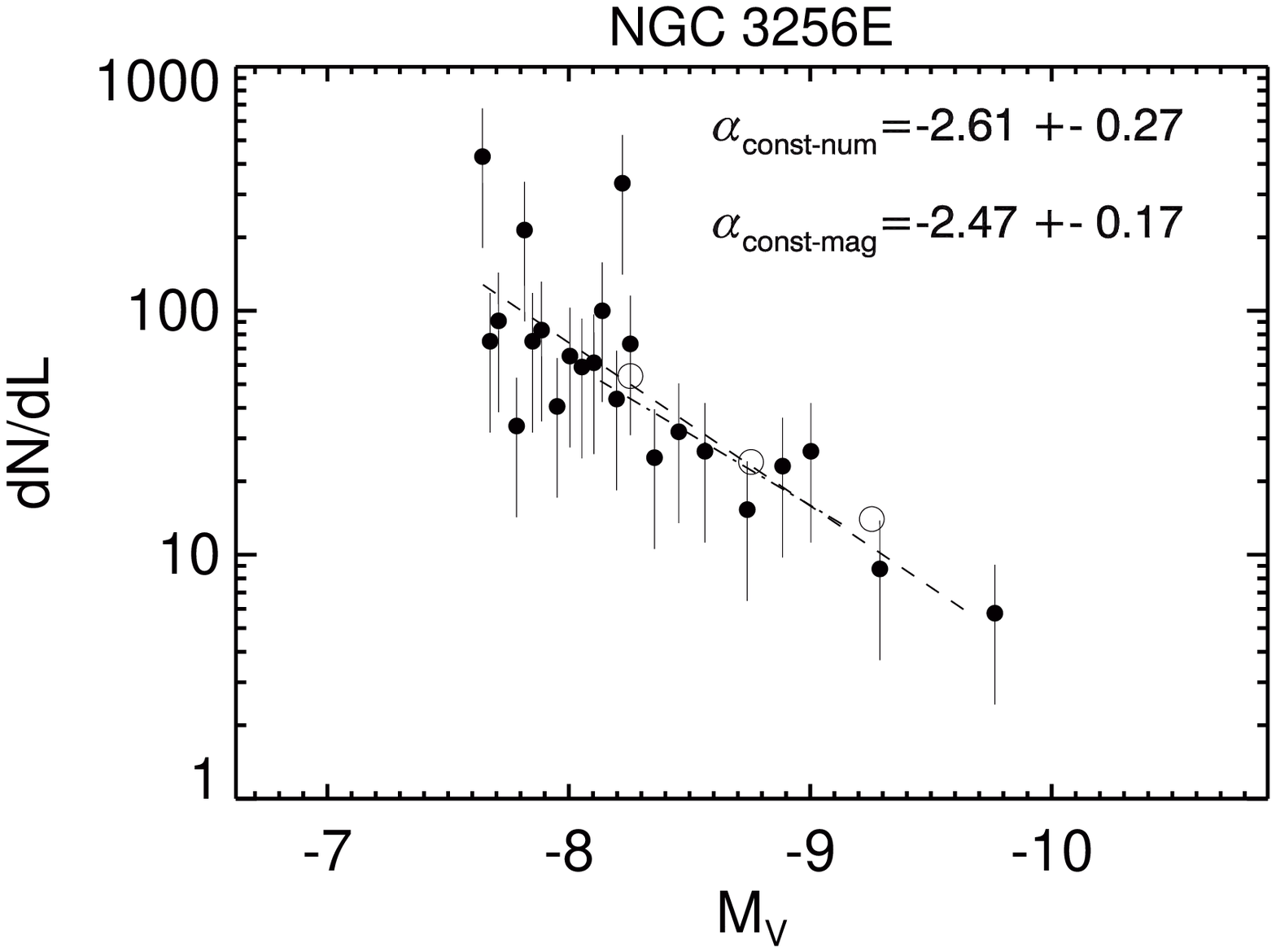}&
\includegraphics[scale=0.45]{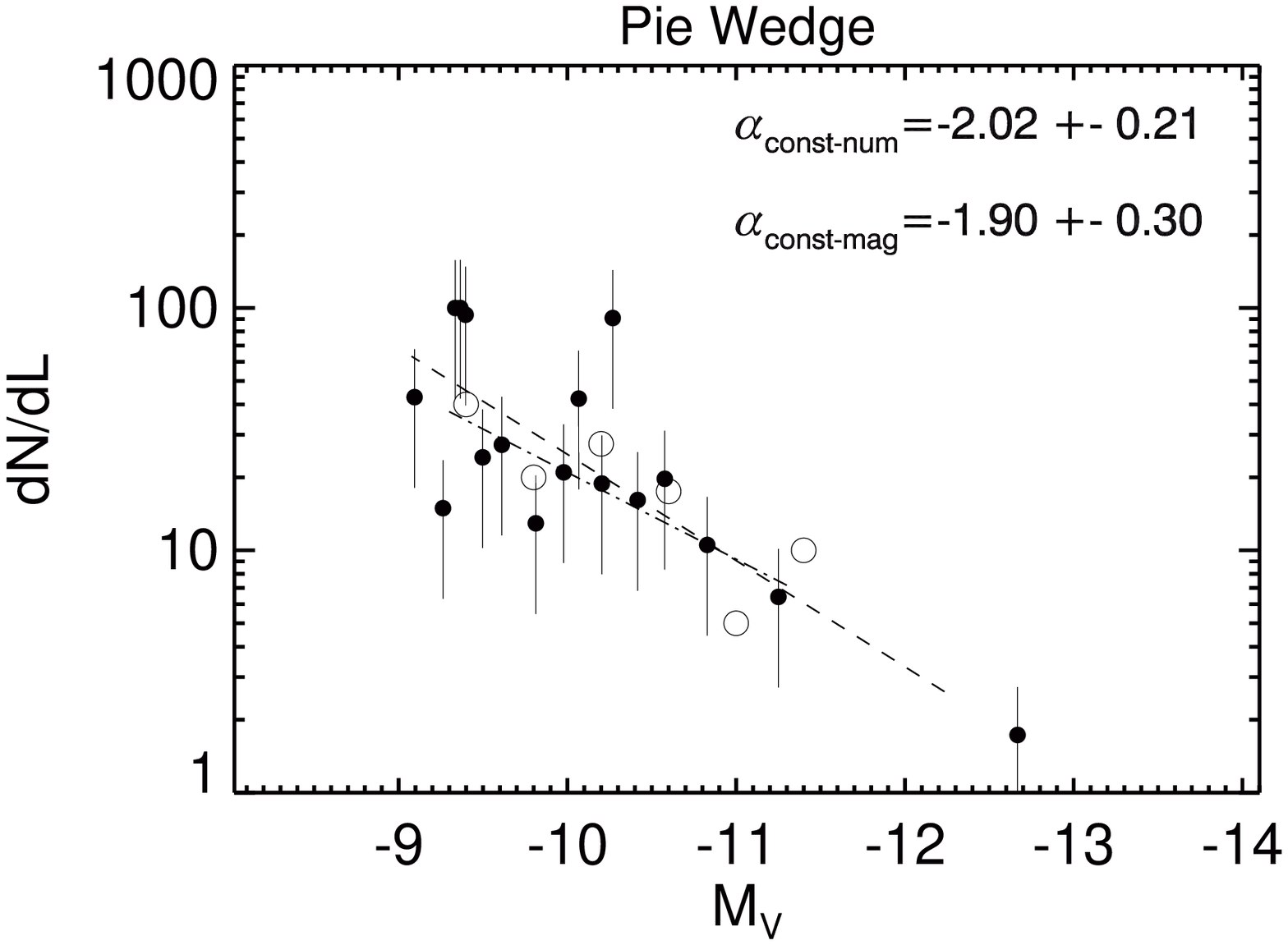}
\end{array}$
\caption{Luminosity functions of clusters in tails where measurements
 were possible.  Filled circles are derived from the constant-number
 method and are best fit by the dashed line (slope given as
 $\alpha_{\rm{const-num}}$; see section 3.2 for details).  Open
 circles fit with a dash-dotted line (slope given as
 $\alpha_{\rm{const-mag}}$) are derived from the constant magnitude
 bin method.}
\label{lum_fun}
\end{figure*}

The LFs of the tail clusters are shown in
Figure \ref{lum_fun}.   These can be described by a power
law, $dN/dL \propto L^{\alpha}$, where $\alpha$ is determined from a
linear fit to log $(dN/d(M_V))$.  We fit our cluster LFs over the $V$
band magnitude range which we
believe to be reasonably complete.  Magnitudes include
  distance moduli from Table \ref{Tab:galaxies} as well as
  aperture corrections,  and they have been dereddened because of foreground
extinction (see Section 3.3.1).  We are unable to measure the LF in NGC
520N and NGC 2623S because they contained too few clusters.

We determine the LFs in two different ways, using both a fixed magnitude bin
width (denoted $\alpha_{\rm{const-mag}}$) and a fixed
number of clusters per bin ($\alpha_{\rm{const-num}}$).  The results
from both methods are
consistent, and we report $\alpha$ values for the latter
because it is better at handling low number statistics.  In all cases,
however, we find that the two values for $\alpha$ are the same, within
the uncertainties.  The typical standard deviation among
$\alpha$ calculated using $B$, $I$, and $V$ band magnitudes is
$\approx$ 0.13, with no systematic trend.  The LFs in
the $B$ and $I$ bands have similar slopes to those in the $V$ band.

In NGC 520S, we find a best fit of $\alpha \sim -2.32
\pm 0.36$ down to $M_V < -8.0$.  We restrict our fit to this
magnitude limit because completeness becomes an issue fainter than
this.  For NGC 2623, we focus on the pie wedge and northern tail.  We
fit the luminosity function in both regions down to $M_V \approx
-9.0$, and find $\alpha = -2.39 \pm 0.34$ for the northern tail and
$\alpha = -2.02 \pm 0.21$ for the pie wedge.  The luminosity function
in NGC 3256E is best
fit with $\alpha = -2.61 \pm 0.27$ down to $M_V = -8.0$.
We find no evidence for a deviation from a simple power-law at the bright end
of the LF in any of our tail regions.

Several studies have shown that the luminosity function can range from
$-1.9 < \alpha < -2.8$ for clusters in a variety of environments
(e.g., Whitmore et al. 2014 and references therein).  We find that our
results for $\alpha$ are within this range.

\subsection{Colors and Ages of Clusters in Tidal Tails}

\begin{figure*}[ht]
\centering$
\begin{array}{cc}
\includegraphics[scale=0.50]{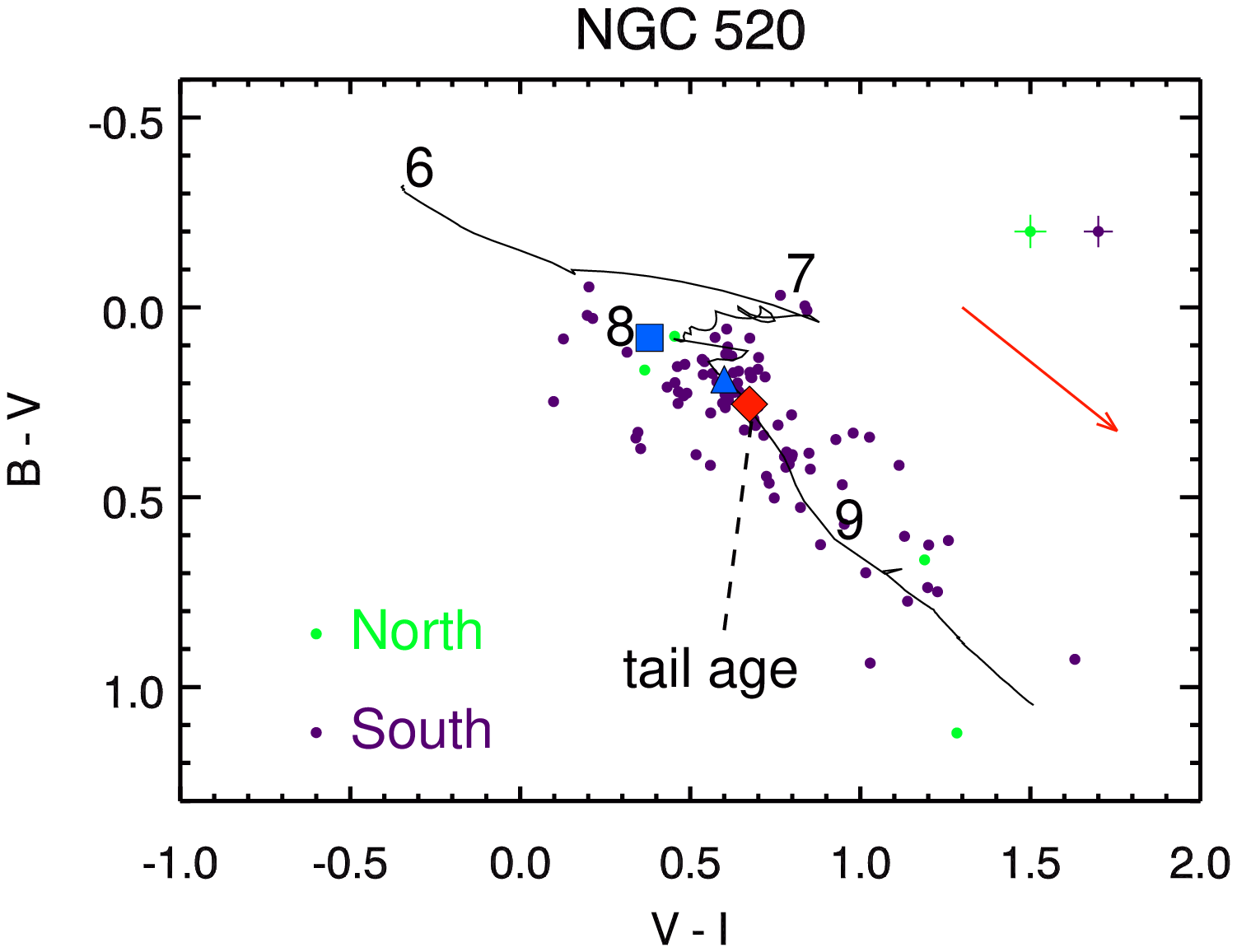}&
\includegraphics[scale=0.50]{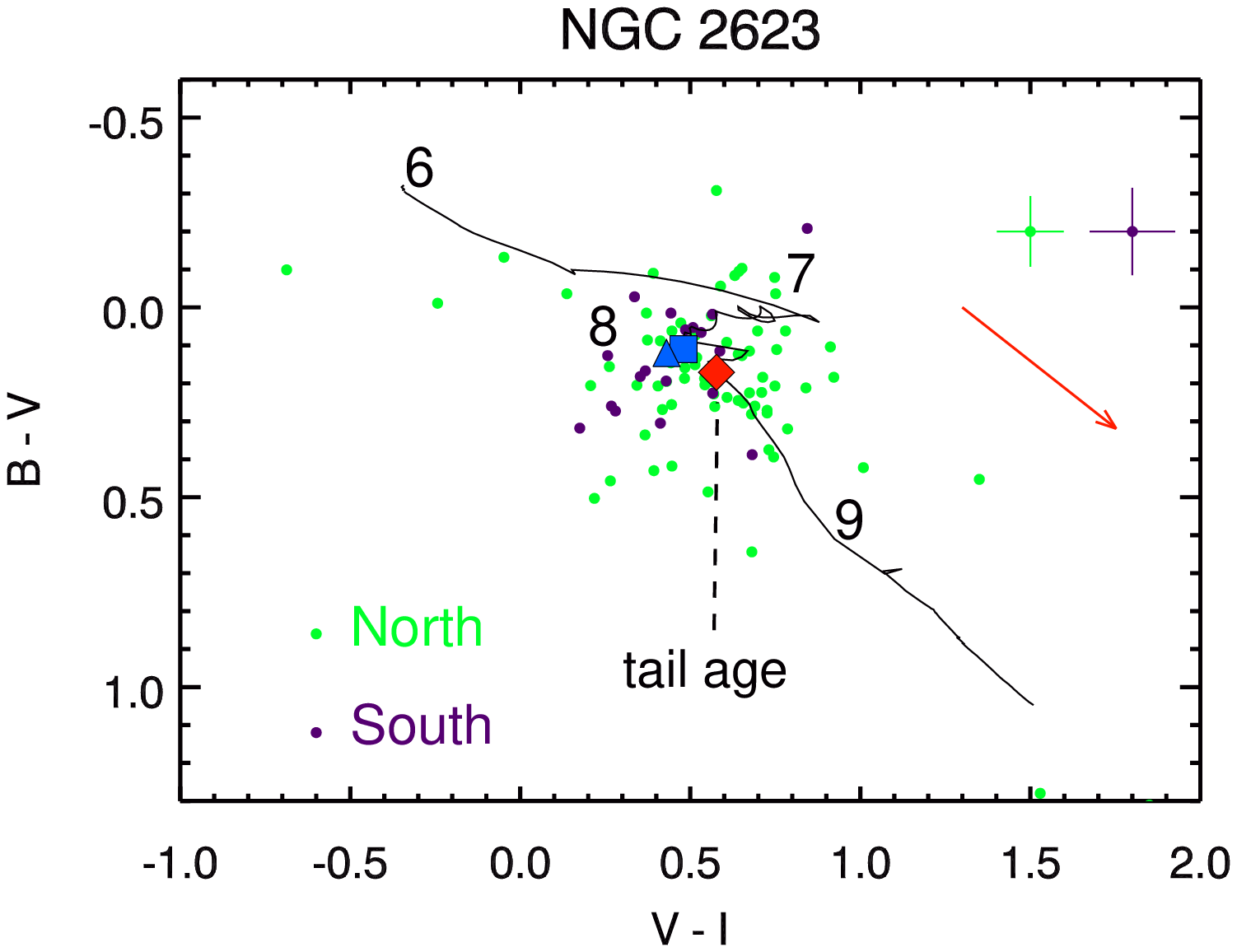}\\
\includegraphics[scale=0.50]{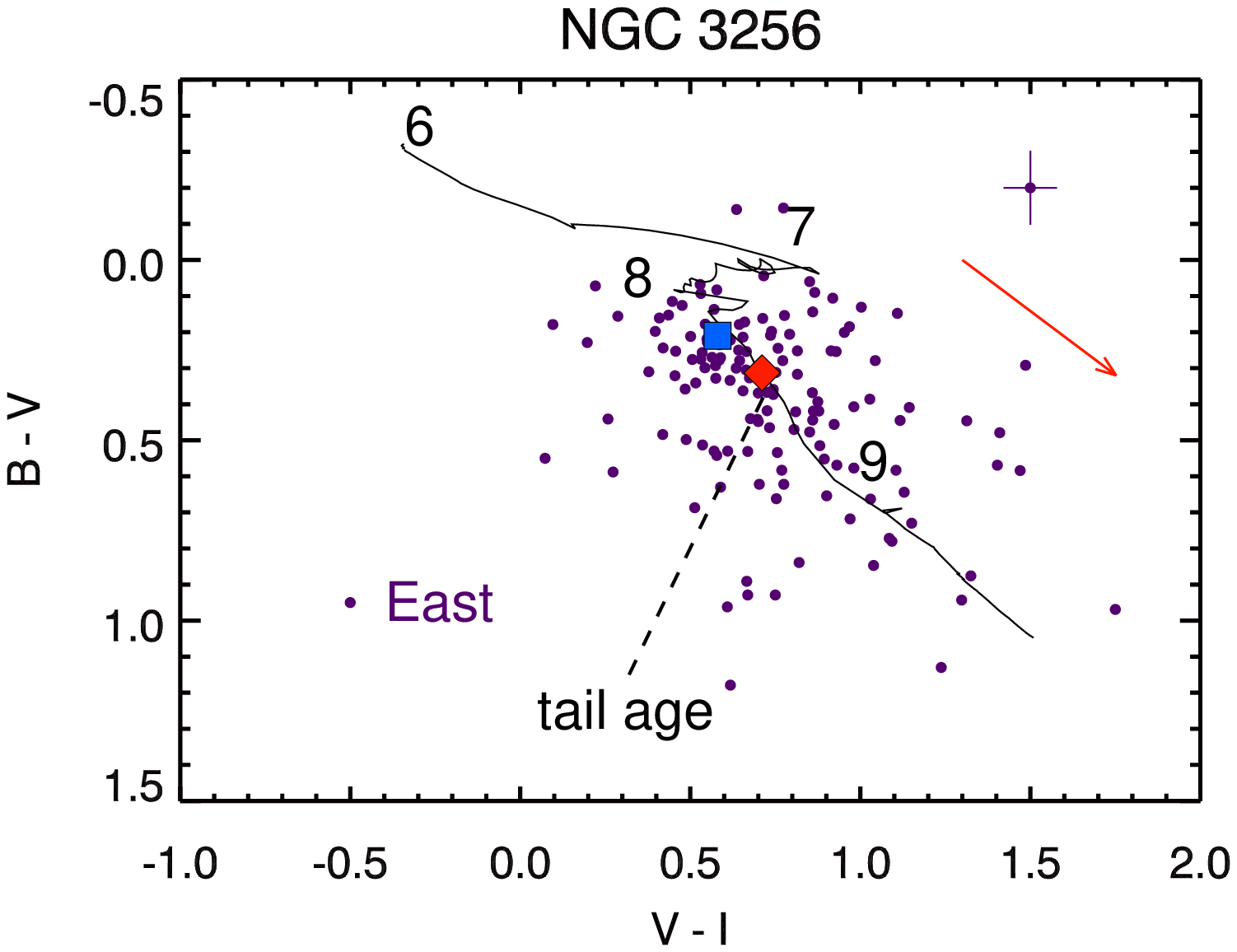}&
\includegraphics[scale=0.50]{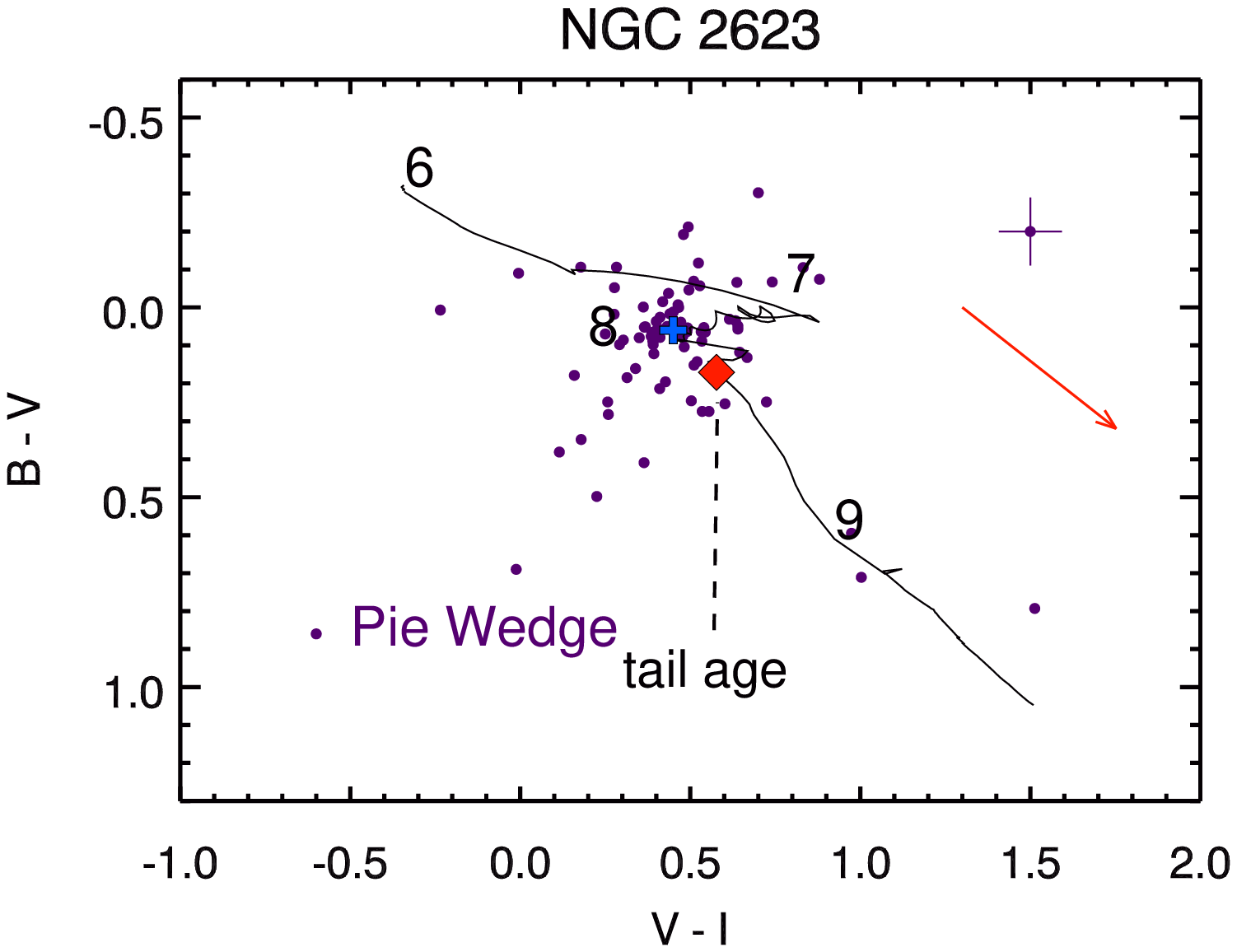}\\
\end{array}$
\caption{Color-color diagrams.  Clusters have been corrected for foreground reddening of the Milky Way by
using the $A_\lambda$ values compiled in Table \ref{Tab:galaxies}.  Solid black lines are BC06
cluster evolution tracks, and numbers represent log ages of clusters.  The reddening
  vectors are shown by the red arrows for $A_V=1$.  For comparison, the
  red diamond marks where the tail age ($\tau_{\rm{tail}}$) would fall on this diagram
  based on the ages shown.  Clusters that formed in the tail are
  younger than the tail and presumably fall above the red diamond on the track.
  The median colors of these young clusters are marked in blue,
  represented by either a square (northern/eastern
  tail), triangle (southern tail), or a cross (pie wedge).
  The points in the upper right-hand corners indicate typical errors
  from photometric uncertainties.}
\label{clus_CCD}
\end{figure*}

We present the color-color diagrams for clusters
in all tails in Figure \ref{clus_CCD}.  The color coding corresponds to different
tails within each merging system.  We also show with blue symbols the median $V - I$ and
$B - V$ colors of clusters that apparently formed after the tidal
tails (we assume tail ages from simulations that are discussed in Section
4.1 and compiled in Table \ref{Tab:clusters}).  The solid line
in each panel in Figure \ref{clus_CCD} is a stellar population model
from G. Bruzual \& S. Charlot (2006, private communication, hereafter
  BC06; and see also
Bruzual \& Charlot 2003) which predicts the evolution of star
clusters from about $10^6$ -- $10^{10}$ yr.
All models assume a metallicity of Z = 0.02, and a
Chabrier (2003) initial mass function.  Numbers mark the logarithmic
age ($\tau$) corresponding to the population.

\begin{figure}[h]
\centering
\includegraphics[scale=0.30]{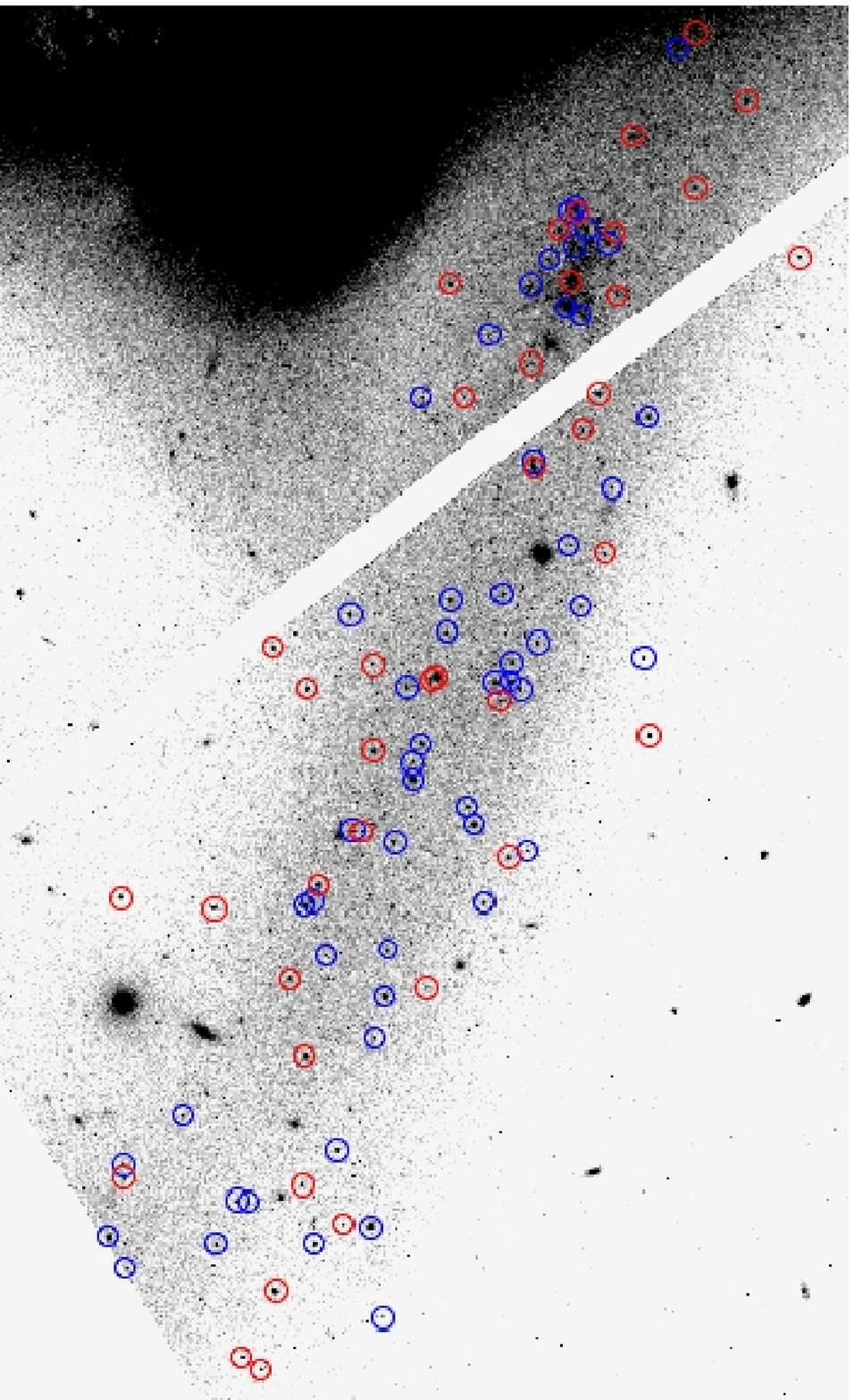}
\caption{Southern tail of NGC 520.  Red and blue circles indicate
  clusters that are redder and bluer than our color cut ($B-V = 0.3$),
  respectively.}
\label{520_ages}
\end{figure}

\begin{figure}[h]
\centering
\includegraphics[scale=0.50]{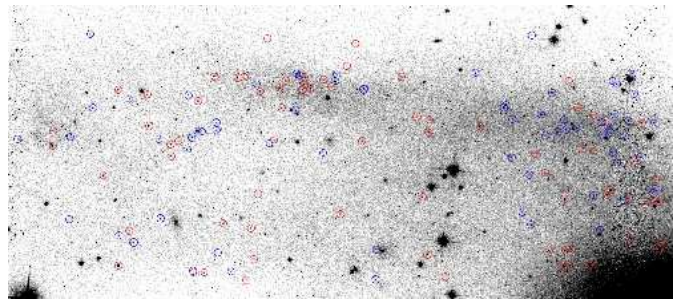}
\caption{Eastern tail of NGC 3256.  Red and blue circles indicate
  clusters that are redder and bluer than our color cut ($B-V = 0.3$),
  respectively.}
\label{3256_ages}
\end{figure}

\subsubsection{The Impact of Internal Reddening on Cluster Colors}

In order to accurately age-date star clusters, we must understand how
reddening affects their measured colors.
We correct each cluster for foreground reddening of the Milky Way by
using $A_\lambda$ values given in the NASA/IPAC Extragalactic
Database, which are compiled in Table
\ref{Tab:galaxies}.  The reddenings were
estimated with the standard relations between them and
the extinction.  A reddening vector for $A_V = 1$ is shown in Figure
\ref{clus_CCD} by the red arrow.  The reddening vector is
nearly parallel to the predicted evolution of cluster colors, making it
challenging to disentangle the effects of age and reddening in the
measured colors.  For example, a young cluster with some extinction will have
the same colors as an older cluster with no reddening.

Figure \ref{clus_CCD} shows that the clusters in each tail have a
measured color spread.  In principle, there are a few reasons this
spread could exist other than photometric uncertainties, such as
internal (and therefore differential) extinction or a range in cluster
ages.  Below, we examine each galaxy individually and determine that
internal extinction does not greatly affect the majority of our
cluster colors.

We first focus on NGC 2623, whose color distributions are shown in the
upper- and lower-right panels of  Figure \ref{clus_CCD}.  The spread
in measured colors is the smallest for any tail studied here, and instead of
spreading along the reddening vector, the cluster colors are
fairly isotropically distributed.  This pattern suggests that
photometric uncertainties are primarily responsible for the color
spread.  We confirm this by creating color-color diagrams of NGC 2623 clusters
in different $M_V$ bins, and while the cluster color spread
increases with fainter magnitudes, the color distribution is
centered near the median colors quoted in Table \ref{Tab:clusters},
regardless of brightness.  If all of the clusters in the tails of NGC 2623 formed at the same
time, clusters obscured by dust would exhibit redder colors than
unobscured clusters, causing a color spread along the
reddening vector. The clusters in both tidal tails of NGC
2623 appear to have formed over a relatively short period of
time.  This is also true in the pie wedge, where the narrow color
ranges are also indicative of coeval cluster formation.

NGC 520 and NGC 3256, on the other hand, both exhibit a spread in
color along the reddening vector.  Here, we use the relative locations
of blue ($B-V < 0.3$) versus red ($B-V \geq 0.3$) clusters, shown in Figures
\ref{520_ages} and \ref{3256_ages}, to assess whether age or reddening are more likely to be
responsible for the observed spread in color.  In the case of NGC 520
(Figure \ref{520_ages}), the bluer clusters tend
to trace out the central, densest portion of the tail, while the redder
clusters are more evenly spread throughout the tail.  Reddening due to
dust would be expected to give the opposite trend, with reddened
clusters mostly near the center of the tail.  Figure
\ref{3256_ages} shows the locations of blue versus red clusters for NGC
3256.  No obvious pattern is observed, but there is no evidence
that red sources are primarily found in the brightest, densest portion
of the tail.  These results support the interpretation
that the larger color spread of the clusters in the tails
of NGC 520 and NGC 3256 is primarily due to a spread of
cluster ages rather than to varying amounts of cluster
reddening.

\begin{figure}[h]
\centering$
\begin{array}{cc}
\includegraphics[scale=0.40]{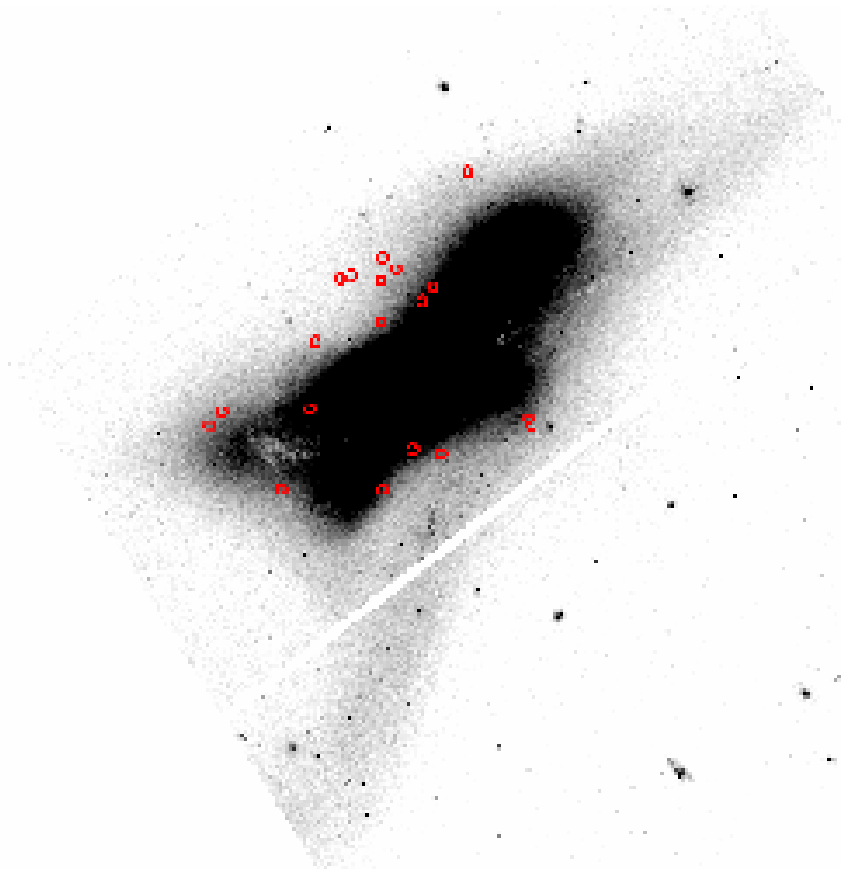}&
\includegraphics[scale=0.60]{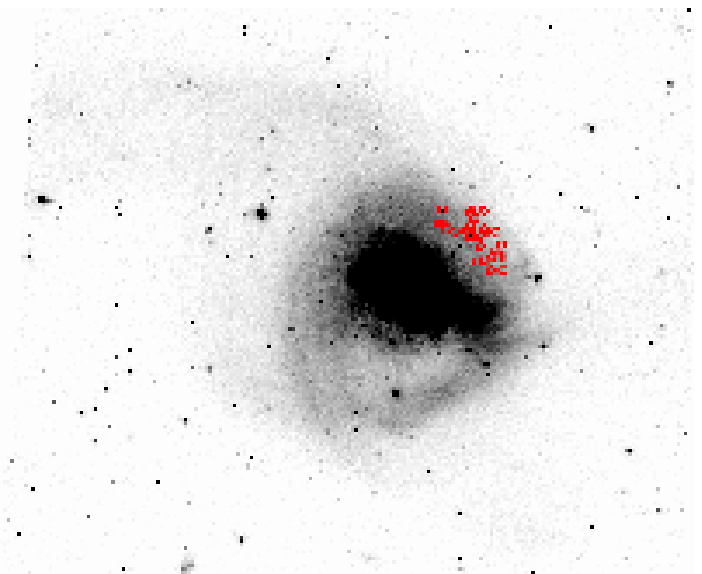}\\
\end{array}$
\caption{Locations of star clusters in the halos of NGC 520 (left) and NGC 3256
  (right).}
\label{halo}
\end{figure}

We now compare the colors of red tail clusters to clusters in the halos of NGC
520 and NGC 3256, which are shown in Figure
\ref{halo}.  Selection and photometry of the halo clusters are
performed using the same methods as in Section 2.  We conclude that
they are fairly similar in color, and
therefore the red tail clusters are likely to be clusters that formed
before the tails but were swept out
during the tail formation.  We will study these halo clusters and
other main body clusters in more detail in a forthcoming paper.  All
of this evidence supports the interpretation that the red sources are
older, unobscured star clusters.

\subsubsection{Ages}

We obtain ages for each cluster candidate by mapping their $B-V$ and $V-I$
colors to the BC06 track and finding the closest match (Figure
  \ref{clus_CCD}).  Ages are fairly accurate in the range $10^6 <
\tau < 10^9$ yr.  Based on the discussion in Section 3.3.1, we assume
no internal reddening.

Figure \ref{clus_CCD} implies that star clusters in the
tails of NGC 520 and NGC 3256 have a broad range of ages up
to $\sim 10^{9.5}$ yr.  The tail ages are marked for
each system with a red diamond, and clusters with colors below the red
diamond are older than the
tail age and might be old disk clusters that predate the merger; we
will hereafter refer to these as ``old'' clusters, and we will refer
to clusters younger than the tail age as ``young'' clusters.  The tail
clusters in NGC 2623 contain ages up to a few $\times 10^8$
yr.  The lack of observed clusters with ages
older than a few $\times 10^8$ yr may simply be because they
are too faint to be observed in our data at the distance of NGC 2623.

\subsection{Diffuse Light in Tidal Tails}

In addition to distinct stellar clusters, we also
study the diffuse stellar light in the tails.  Because
the tidal tails form from material stripped from the native disks, this
diffuse light should consist of a mixture of old
disk stars, faint newly formed stars, and dissolved clusters.  While
individual stars are unresolved, the integrated colors of the diffuse
light can help to constrain the age of the stellar populations (e.g.,
Gallagher et al. 2010).  The light will be dominated by the brightest and youngest unresolved
objects, so we expect that our age estimates will be lower limits.

\begin{figure*}[ht]
\centering$
\begin{array}{cc}
\includegraphics[scale=0.50]{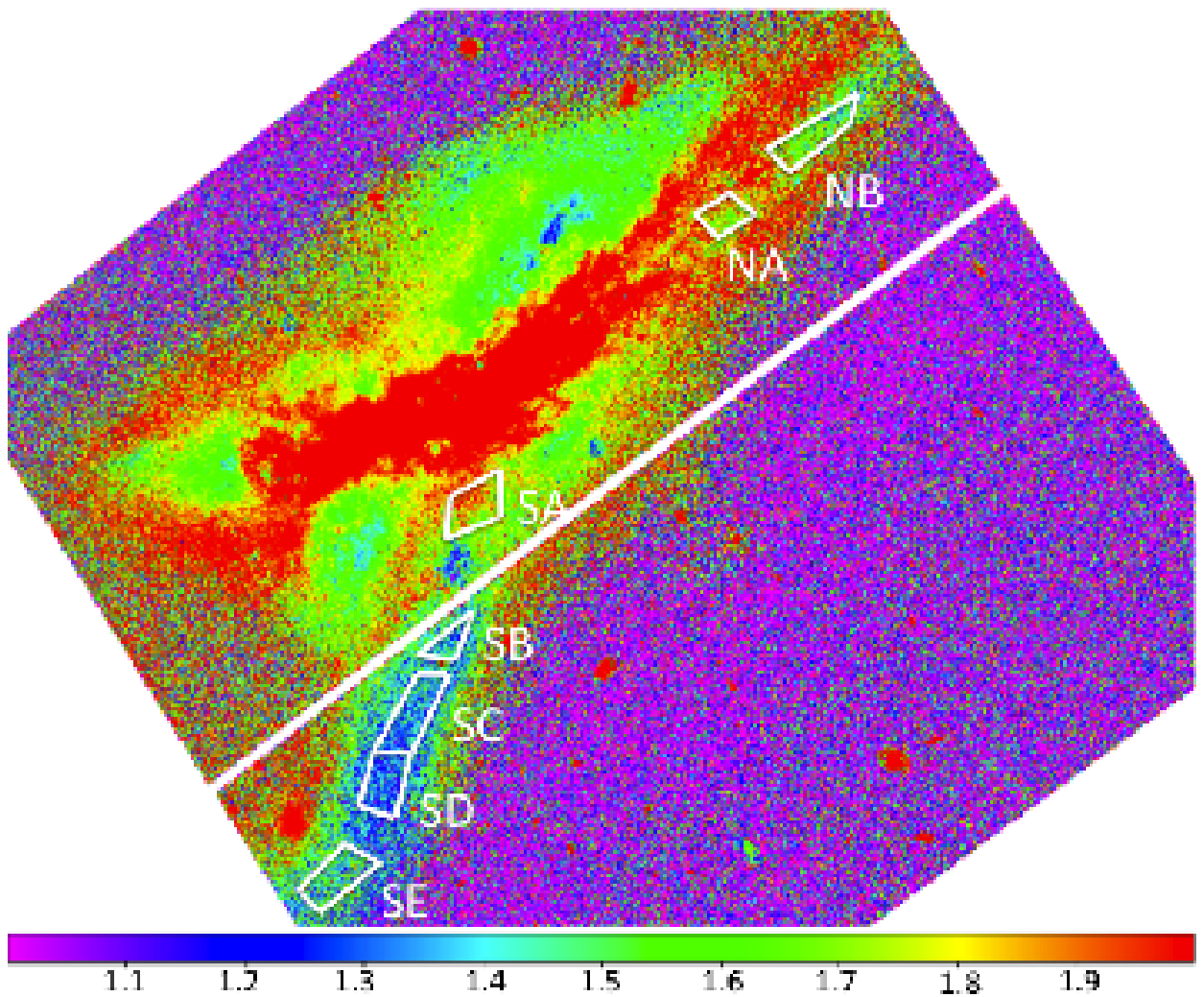}&
\includegraphics[scale=0.50]{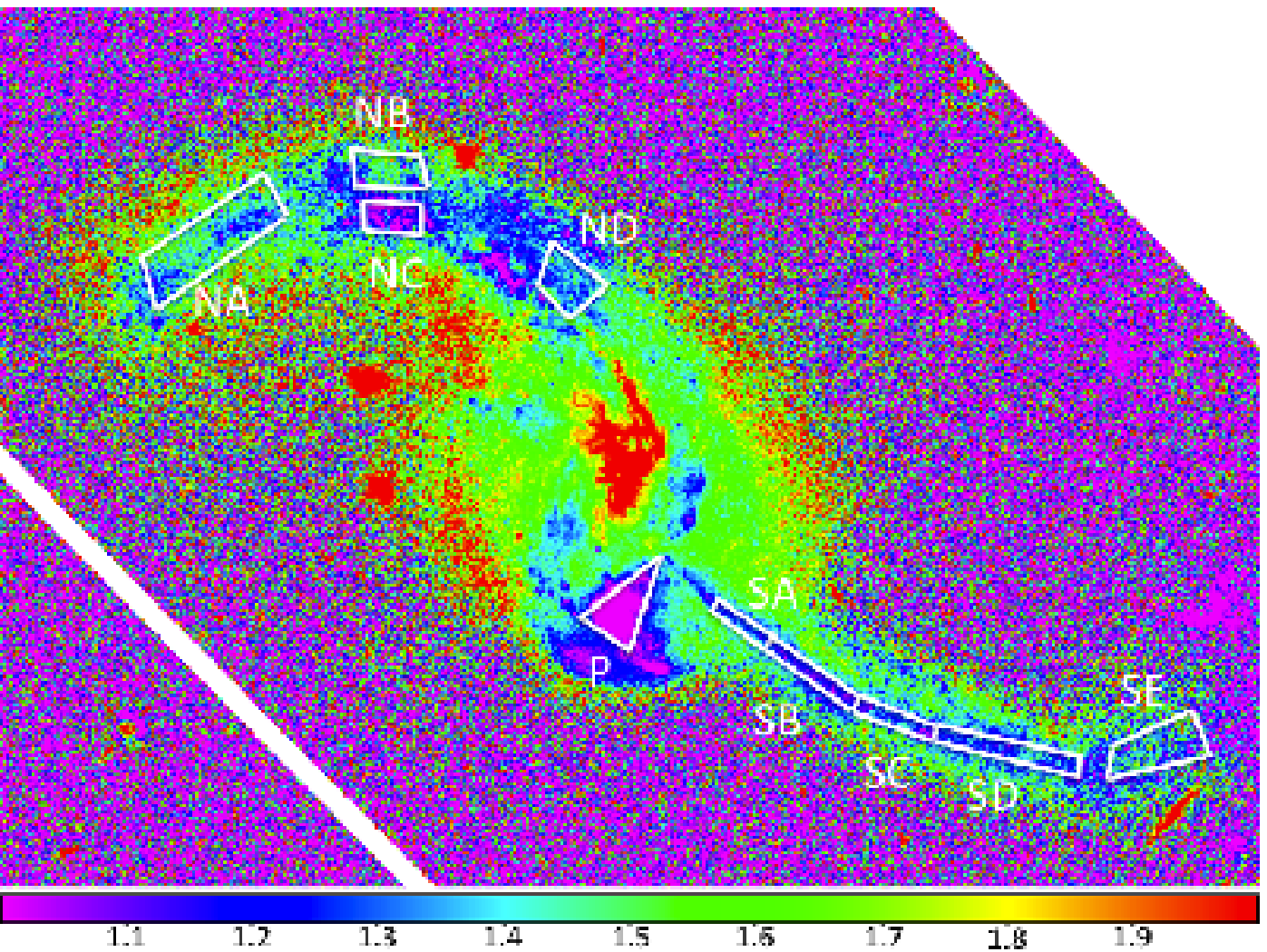}\\
\includegraphics[scale=0.50]{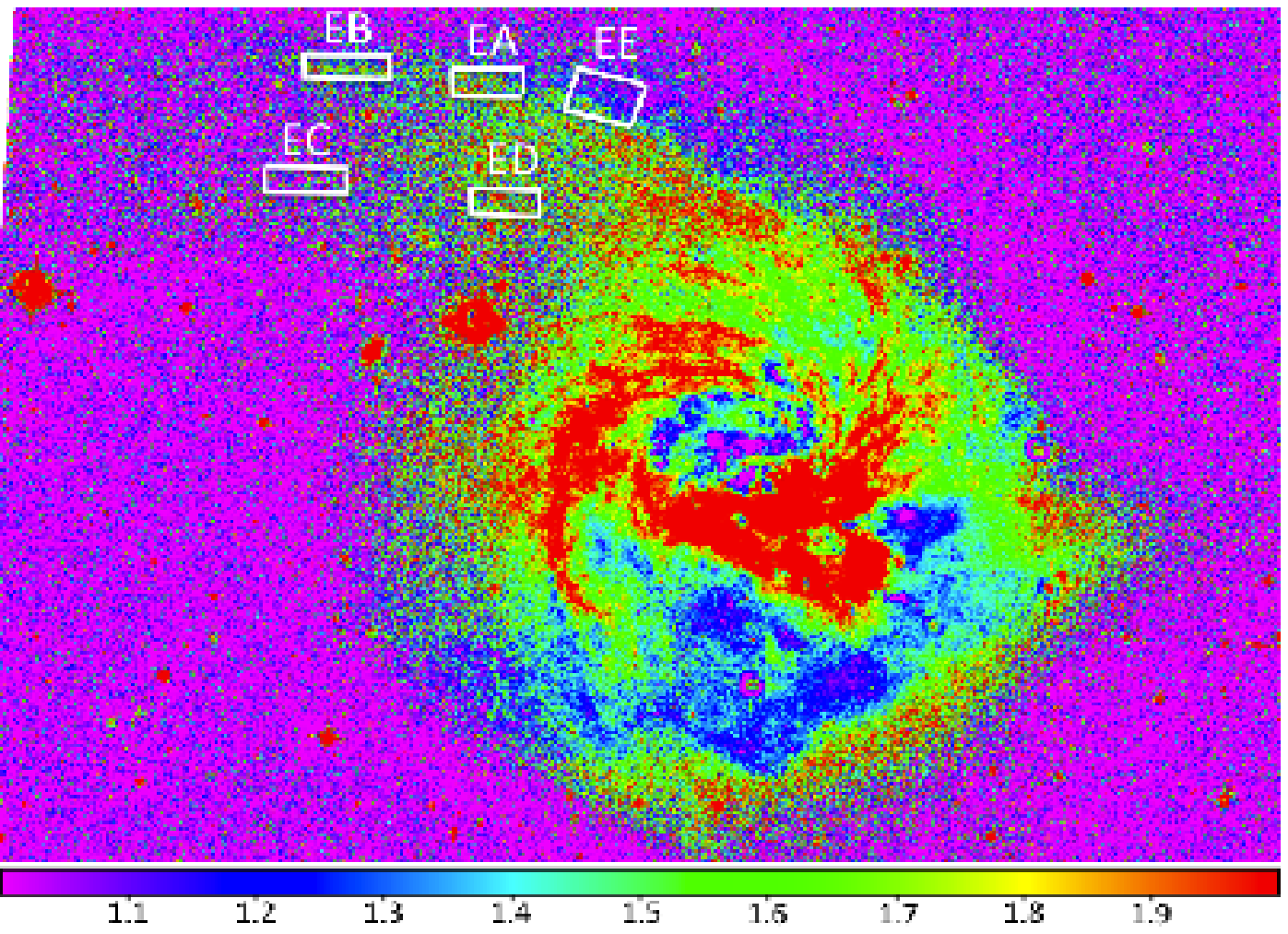}
\end{array}$
\caption{$B-I$ images of NGC 520 (top left), NGC 2623 (top right),
 and NGC 3256 (bottom left).  The scale on bottom gives the $B-I$
 color values.}
\label{diffuse}
\end{figure*}

We create color images: $B-V$, $V-I$, and $B-I$ for all three
galaxies, and we show the $B-I$ images in Figure \ref{diffuse}.  The
images were boxcar-smoothed with a 4 pixel kernel width.  The main bodies of the
mergers are very red because they contain large
amounts of dust.  The tidal tails are easily seen in Figure
\ref{diffuse} as bluer regions where internal extinction is low
(except, perhaps, in NGC 520N, as also suggested by our analysis of
clusters in Section 3.3.1).  We therefore assume that colors in
the tail regions are not greatly affected by internal extinction, and
we believe diffuse light colors trace the ages of stellar populations
relatively well.

\begin{figure}[h]
\centering
\includegraphics[scale=0.50]{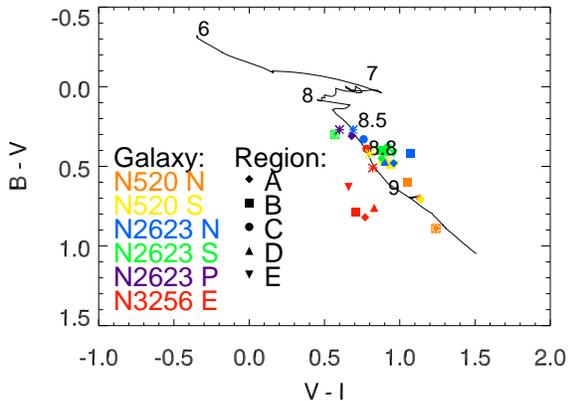}
\caption{Diffuse light colors for all three galaxies.  Regions indicated
  in the legend correspond to Figure \ref{diffuse}.  Asterisks mark
  the median color of clusters found in that tidal tail that presumably predate the
  merger.  Asterisks inside squares indicate that there were very few
  clusters that predate the merger ($ \leq 3$).  The solid black line
  is a BC06 cluster evolution track.}
\label{diffuse_CCD}
\end{figure}
 
We measure the $V - I$ and $B - V$ colors in different tail regions
(white boxes in Figure \ref{diffuse} labeled from ``A'' up to ``E'' for each tail) after
subtracting the point sources.  Colors are calculated from the
  ratio of fluxes on a pixel by pixel basis and corrected
  for differences in instrumental zeropoints and foreground
  extinction.  The diffuse light regions were chosen
to cover most of the optically bright portions of the tails while
avoiding breaks in field coverage and saturated
foreground stars. The values in Table \ref{Tab:diffuse} are the mean
color for each region.  Figure \ref{diffuse_CCD} plots these
values on a color-color diagram overlaid with the BC06 track.
Figure \ref{diffuse} reveals that the diffuse stellar light in all tidal tails in our sample
exhibits a common pattern (with the exception of NGC 3256E).  While the
tidal tails are fairly blue ($B - I \leq 1.4$), we find that there is
a gradient with $B-I$ color; it is redder near the edges of the tail
and bluer near its axis.  As
previously discussed, this is not likely a reddening effect, so we
believe that we are observing a gradient in the mean ages of the stars
across these tails.  In Section 3.3.1 we discussed the
tendency for younger clusters to be near the center of the tail
axis, leaving mostly older clusters along the tail edges,
especially in NGC 520S.  The diffuse light images further support a
lack of young stars near the edges of the tails.  We believe this to
be the first time such a gradient \emph{across} the tails has been
observed in these galaxies.

\begin{table}
\caption{Diffuse Light Measurements}
\label{Tab:diffuse}
\begin{center}
\begin{tabular}{lccc}
\hline\hline						
Galaxy & Region & $\langle$V - I$\rangle_{\rm{diffuse}}$ & $\langle$B - V$\rangle_{\rm{diffuse}}$\\
\hline
NGC 520  	& NA & 1.13 & 0.70\\
  	& NB & 1.05 & 0.60\\
 	& SA & 1.14 & 0.71\\
 	& SB & 0.94 & 0.49\\
 	& SC & 0.90 & 0.47\\
 	& SD & 0.88 & 0.46\\
 	& SE & 0.97 & 0.49\\
\hline
NGC 2623	& NA & 0.96 & 0.48\\
	& NB & 1.07 & 0.42\\
	& NC & 0.76 & 0.33\\
	& ND & 0.90 & 0.47\\
	& SA & 0.88 & 0.45\\
	& SB & 0.87 & 0.40\\
	& SC & 0.92 & 0.38\\
	& SD & 0.91 & 0.41\\
	& SE & 0.96 & 0.42\\
        & P & 0.68 & 0.31\\
\hline
NGC 3256	& EA & 0.77 & 0.82\\ 
	& EB & 0.71 & 0.79\\
	& EC & 0.78 & 0.39\\
	& ED & 0.83 & 0.76\\
	& EE & 0.66 & 0.63\\
\hline
\end{tabular}
\end{center}
\end{table}

Figure \ref{diffuse_CCD} reveals that the northern tail of NGC 520 has the
reddest diffuse light of any tail in our sample.  This is likely due to dust, as
the red $B-I$ color in Figure \ref{diffuse} extends from the main body
straight through the tail.  Region A in the southern tail is also almost
certainly affected by dust.  We believe that the colors of the diffuse
light in the other regions of the southern tail are dominated by ages
of the underlying populations and not extinction because of the
absence of dust lanes in this tail.  These regions have
an age $\tau$ of around 800 Myr.   

In NGC 2623, most of the regions in the north tail have colors consistent with
stellar populations of $\sim 700$ Myr old.  Region NC, however, is bluer
and yields an age of $\sim 500$ Myr.  We note that the diffuse light
in the pie wedge (region P) has $B - V$ and $V - I$ colors close to
region NC, as seen in Table \ref{Tab:diffuse}, which
might provide clues to its formation, as we discuss further in Section
4.1.2.

Figure \ref{diffuse_CCD} shows that the diffuse light in NGC 3256 is
perhaps the most peculiar of our sample.  Only region EC falls near
the BC06 track, with an associated age $\tau \approx 650$ Myr. This
would indicate that the diffuse light population in region EC is
similar to the majority of the diffuse populations in the tails of NGC
2623.  It would be plausible that region EC falls on a local region of
enhanced star formation, causing its diffuse light colors to be bluer
than other regions of NGC 3256E.
However, region EC does not include any young star clusters,
instead containing only one cluster whose colors suggest that it predates
the merger.  The rest of the regions we measured are gathered around $V-I
\approx 0.7$ and $B-V \approx 0.8$ and are off the BC06 track.  Regardless,
these offset colors suggest faint population ages of
about 800 -- 1000 Myr.  It is likely
that most of these diffuse light regions do not have colors consistent
with the BC06 track because of the low surface brightness of NGC
3256E, resulting in large photometric errors.  It is unlikely
that H$\alpha$ line emission could be responsible for the offset from
the model, although this emission would move measurements in the
observed direction relative to the models, because such recent star
formation is usually quite bright and lumpy.

\section{DISCUSSION}

\subsection{Comparing Cluster Ages with Tail Ages}

The age of the tidal tails, which we define as $\tau_{\rm{tail}}$, refers
to the time since perigalacticon of the two progenitor galaxies,
because they form from material ejected after the galaxies
undergo their closest approach.  The ages of tails can be
estimated from simulations and are probably accurate to within $\sim
20\%$ ($\sim 30 -100$ Myr; Barnes \& Hibbard 2009).  In this section,
we briefly summarize previous simulations of our sample of mergers,
and we compare tail ages from those simulations with our measured
cluster ages.

\begin{figure*}[ht]
\centering$
\begin{array}{cc}
\includegraphics[scale=0.45]{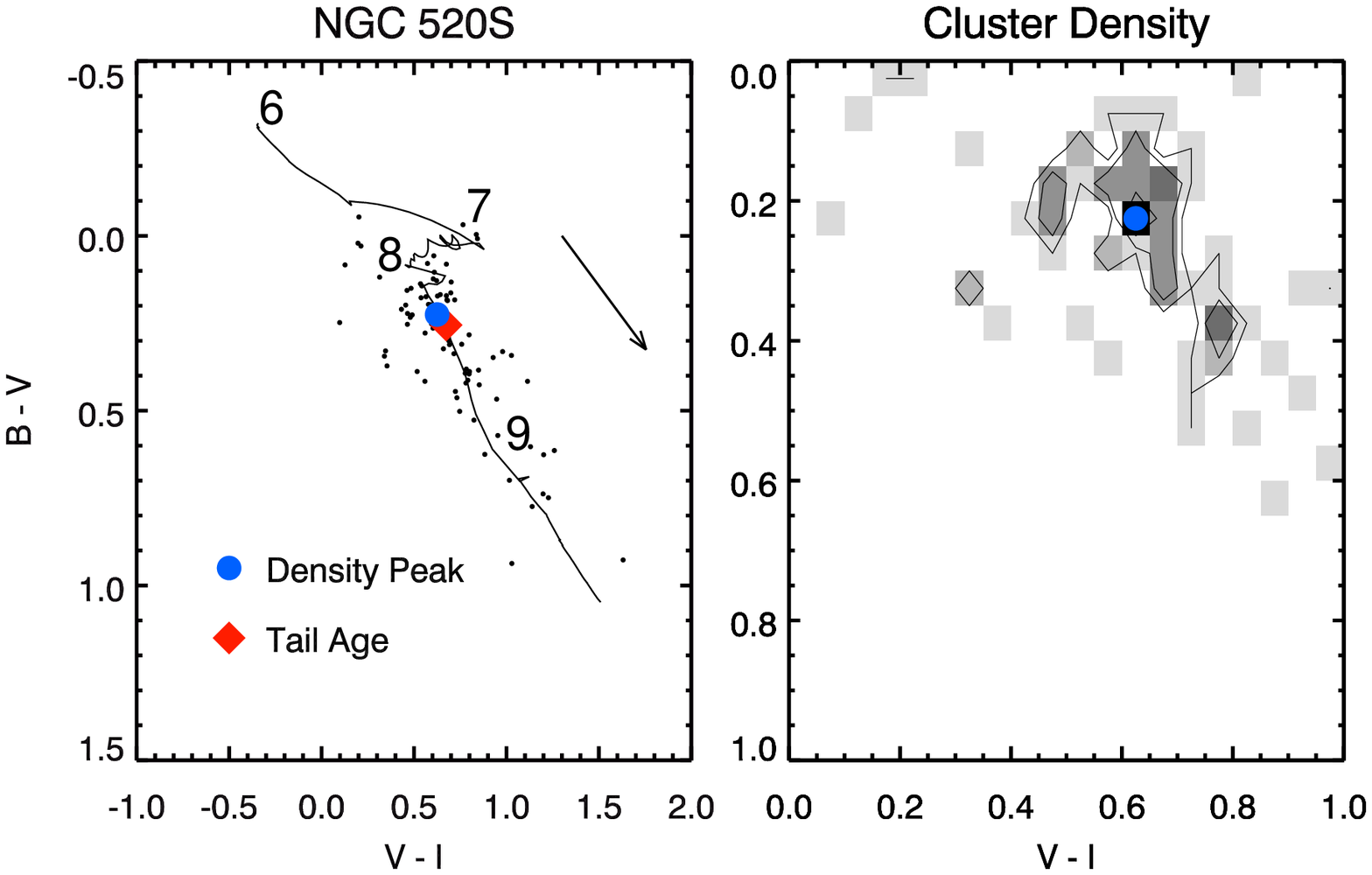}&
\includegraphics[scale=0.45]{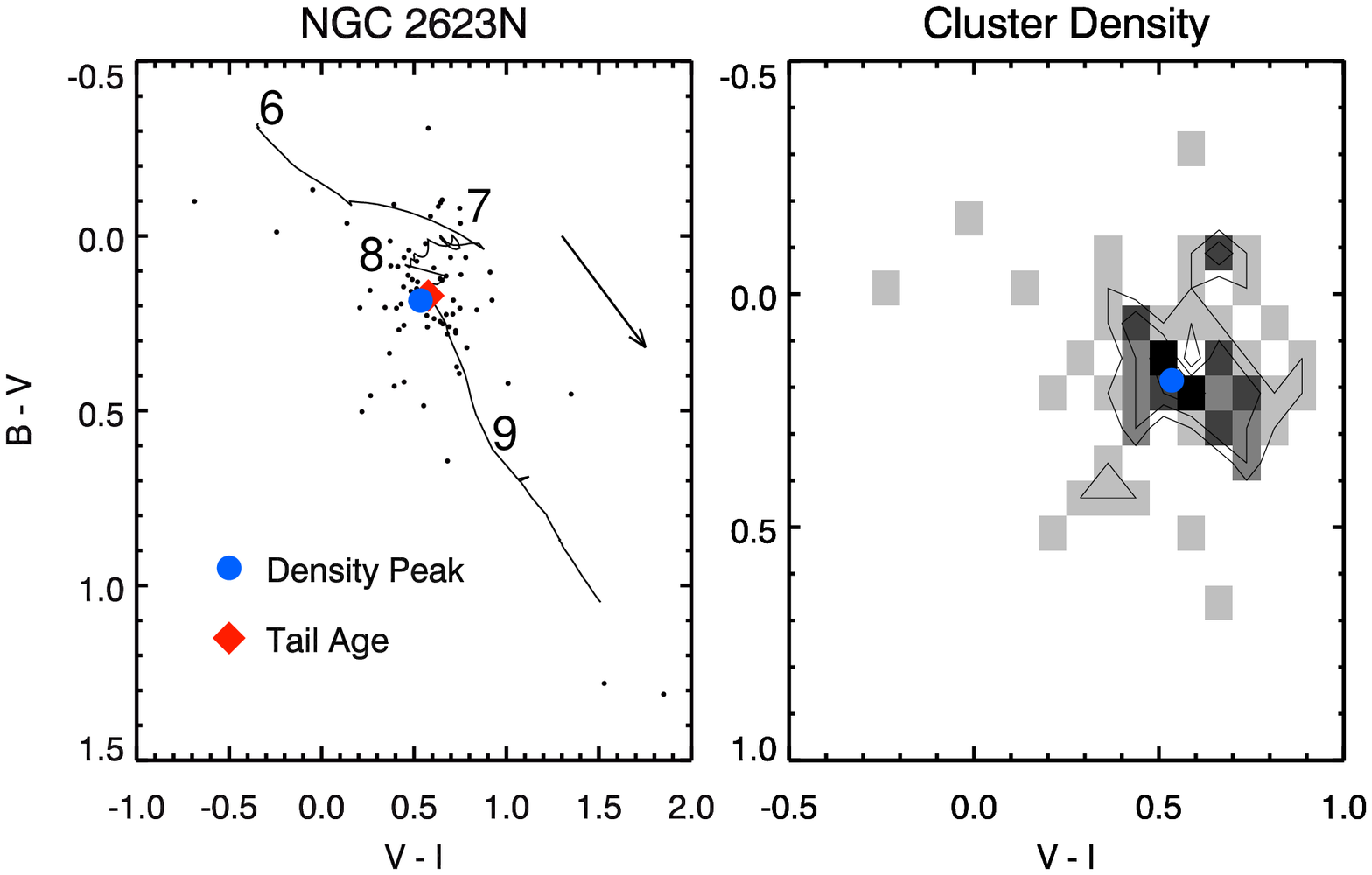}\\
\includegraphics[scale=0.45]{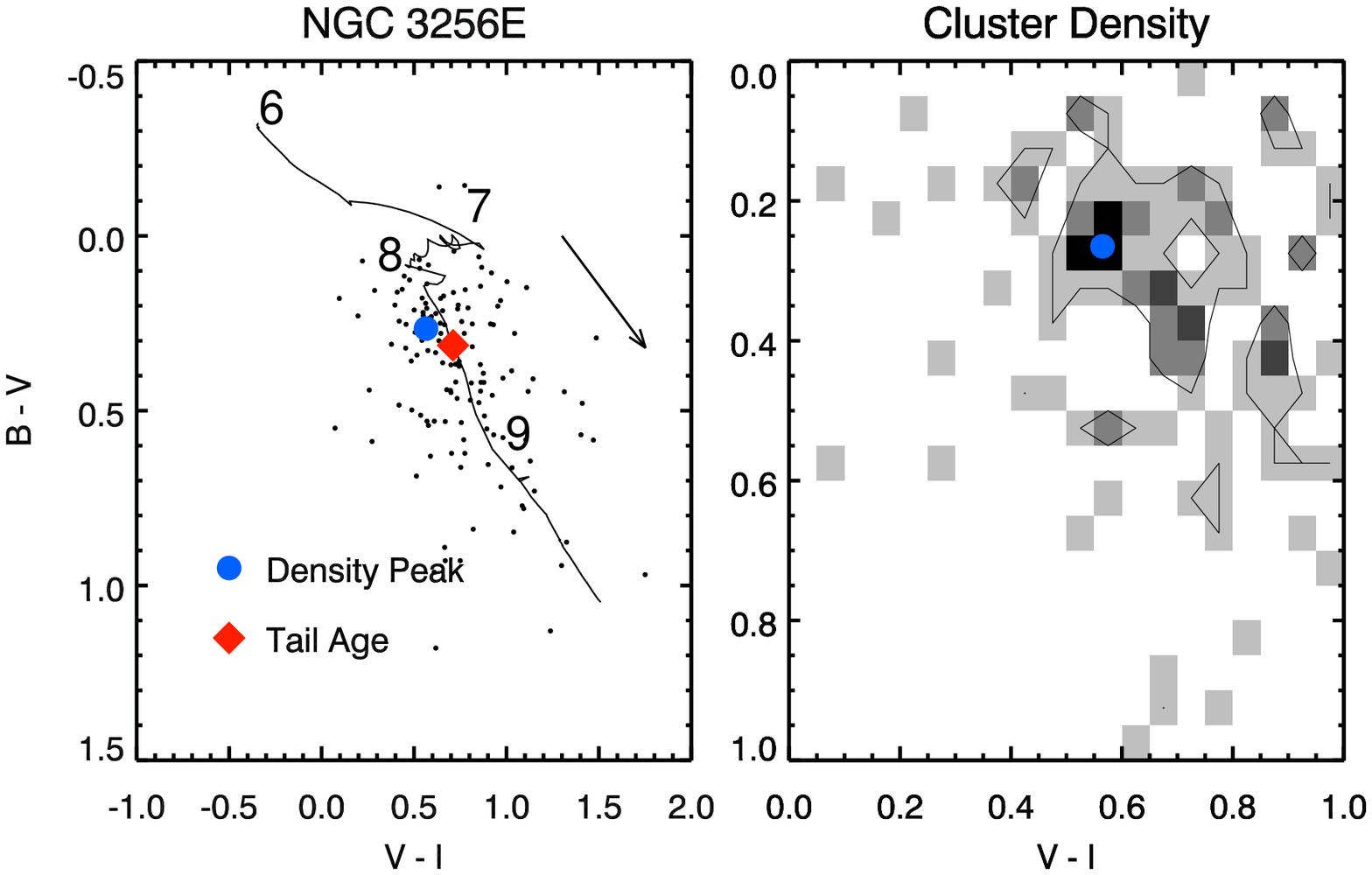}&
\includegraphics[scale=0.45]{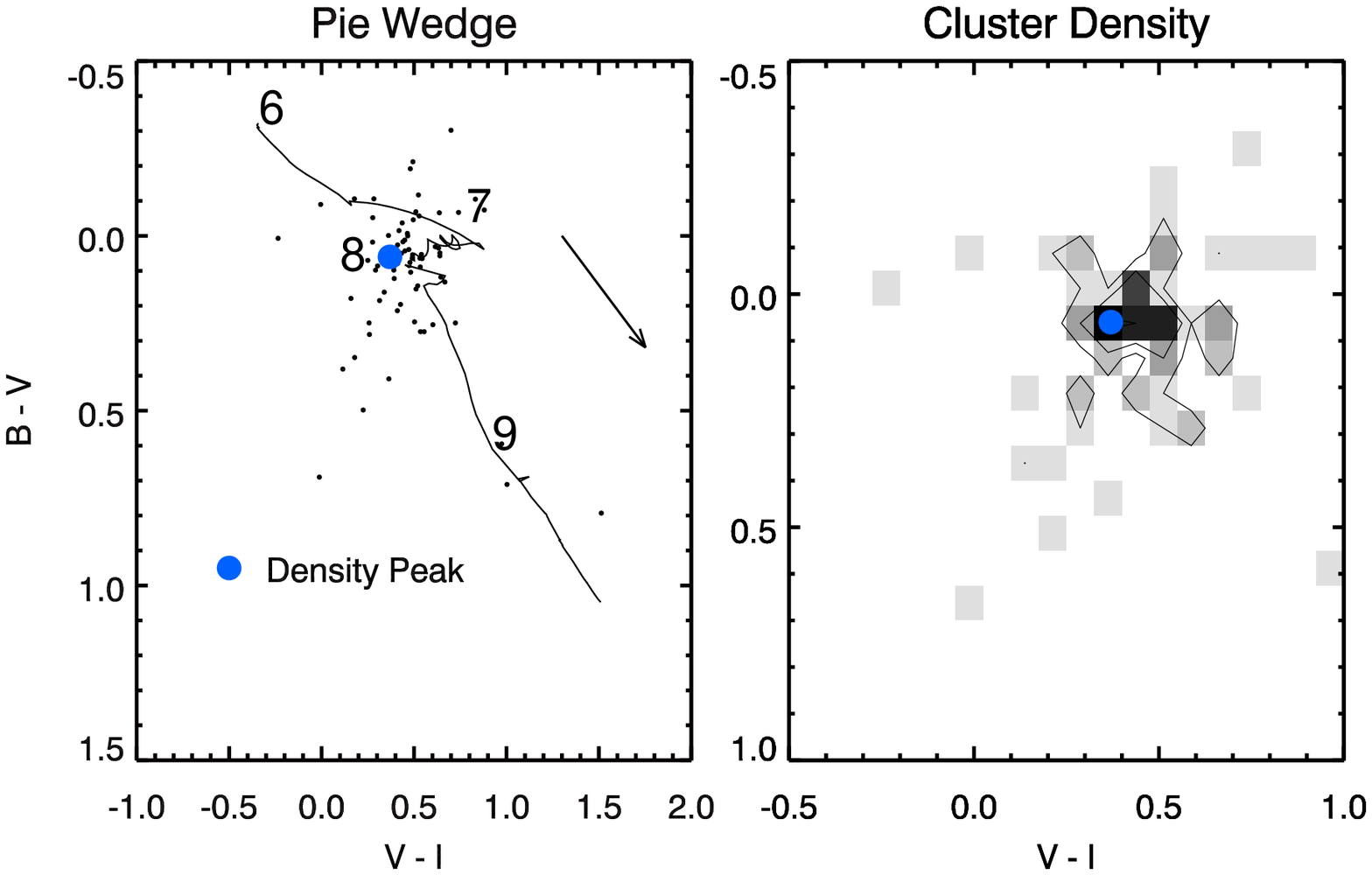}\\
\end{array}$
\caption{Color-color diagrams used to quantify when the peak cluster
  formation occurred in each system where possible.  Right panels show
  densities of clusters in color-color space, with contours
  overplotted.  The peak is marked with a blue circle.  Left panels
  are similar to Figure \ref{clus_CCD}, with density peaks from the
  right panels included.}
\label{dens_CCD}
\end{figure*}

We find that, for most systems, there is a period of increased cluster
formation near the time that the tidal tails formed.  We
quantify the time of this cluster formation (hereafter denoted $\tau_{\rm{clus}}$) by
comparing the density peak of the color distribution with cluster evolution model
predictions.  These density plots are shown in the right panels of
Figure \ref{dens_CCD}.  Contours are plotted, and the density peak is
marked with a circle.  The colors of the peak are then compared with a BC06 track
to obtain an age (shown in left panels of Figure \ref{dens_CCD}).

\subsubsection{NGC 520}

The interaction history of NGC 520 is the least well
understood of our sample.  Stanford \& Balcells (1991) ran the most recent
simulation of the system, comparing to infrared and
optical images, as well the \ion{H}{1} velocity field.  Their best match to these
observations are at a simulated age of $\sim$ 300 Myr ago.  The system is particularly
complicated, however, because Stanford \& Balcells find that their simulations
were not fully consistent with two colliding disk galaxies, due
to the presence of a third dwarf galaxy, UGC 957.  They conclude that
the southern tail formed from the interaction of two massive disk
galaxies and that the northern tail is the result of a passage by UGC
957.  If the two tails formed via different interactions, we might
expect differences in their cluster ages.  Despite these complexities,
we adopt a best guess of $\tau_{\rm{tail}} \sim 300$ Myr.

The northern tail of NGC 520 contains only three young
clusters, making a color density plot impractical here.
However, the three clusters are tightly grouped around 100 Myr
on the BC06 track; we therefore approximate their ages to be $\tau_{\rm{clus}}
\sim 100$ Myr.  The majority of the south
tail clusters appear to be older, and, based on their color
distribution in Figure \ref{dens_CCD}, they are consistent with a period of
cluster formation that began around $\tau_{\rm{clus}} \sim 260$ Myr ago.

The differences in the numbers and ages of the clusters detected in
the north and south tails appear to be consistent with the scenario
put forth by Stanford \& Balcells.  The passage of a dwarf galaxy
would not necessarily result in a starburst as
intense as one resulting from two equal-mass galaxies, because the
dwarf galaxy would strip less gas to form clusters.  We detect many
more clusters in the southern tail, which Stanford \& Balcells predict
is the result of a disk-disk interaction, than the northern tail,
which likely formed from the passage of the dwarf galaxy UGC 957.
Stanford \& Balcells speculate that the dwarf galaxy might have
formed the northern tail as the disks were interacting, around the
same time the southern tail was forming.  However, the tight age grouping of
the three clusters in NGC 520N seems to indicate that cluster formation took
place earlier than in NGC 520S.  

\subsubsection{NGC 2623}

Privon et al. (2013) simulated NGC 2623 and estimated that an age
$\tau_{\rm{tail}} = 220 \pm 30$ Myr best reproduces spatial and
kinematic distributions of \ion{H}{1}.  Privon et al. also
successfully recreated the pie wedge region, showing that it is probably material from the northern tail that has
fallen back through the main body and is currently on a southward
trajectory.

We measure clusters in NGC 2623N to have a peak density in color
corresponding to an age of $\tau_{\rm{clus}} \sim 230$ Myr; this supports the
scenario that cluster formation began at the same time that the tail formed.
There are some clusters that appear to be older than the tail,
however, as we discussed in Section 3.3.1, we believe this is due to
photometric scatter resulting from
low S/N.  NGC 2623S possesses too few clusters to reliably measure $\tau_{\rm{clus}}$,
although it is clear from Figure \ref{clus_CCD} that most of the
clusters are slightly bluer than the color associated with the tail
age.  It is possible that formation of clusters in the south tail was
somewhat delayed relative to the tail itself, but uncertainties in
tail and cluster ages are too large to definitively establish such a delay.

Inspection of Figures \ref{clus_CCD} and \ref{dens_CCD} shows that
clusters in the pie wedge have a very narrow range of colors,
suggesting that they all formed at approximately the same time; we measure 
$\tau_{\rm{clus}}$ to be $\sim 100$ Myr.  These clusters are clearly bluer
and hence younger than clusters in the tails of NGC 2623. This result
is consistent with the scenario proposed by Evans et al. (2008), where the
pie wedge formed from debris from the inner region of the northern
tail.  They suggest that the debris fell back through the main body, inducing a burst of
star formation (see also Privon et al. 2013).  While Privon et al. do not
provide an age for the pie wedge, their Figure 8 shows snapshots of the
merger at different times.  At 85 Myr ago, the debris can be seen
falling southward, past the main body, roughly consistent with our
cluster formation timescale.

\subsubsection{NGC 3256}

Recent unpublished results from simulations place NGC 3256 at
$\tau_{\rm{tail}} = 450 \pm 50$ Myr (G. Privon, private
communication).  The simulation is based on spatial and kinematic
\ion{H}{1} images of the system and was run using the same methodology
as for NGC 2623.  

Clusters in the eastern tail have the largest spread in color of any
system in our sample, with a weak density peak, $\tau_{\rm{clus}} \sim
260$ Myr.  This suggests a longer duration of cluster formation in
this tail.  These ages are consistent with previous results.  Trancho et
al. (2007) obtained spectroscopic ages for three clusters in
the western tail, and found two clusters with ages of $\sim$ 80
Myr, and a third with an estimated age of $\sim$ 230 Myr.  The
latter fits well within the range that we find for clusters in the
eastern tail.

\subsubsection{The Timescale of Cluster Formation}

In the three systems studied here, we have found that cluster
formation typically begins with the tidal tail formation, although in
a few cases a delay cannot be ruled out.  This type of delay was also
found by Bastian et al. (2005) in NGC 6872.  We therefore conclude
that, in general, $\tau_{\rm{clus}} \lesssim \tau_{\rm{tail}}$.

$\tau < 10$ Myr clusters have not formed in any of the tails studied
here.  We confirm this by examining \emph{HST} H$\alpha$ images of the
two most recent mergers, NGC 2623 and NGC 520, and find no H$\alpha$
emission in any of their tails.  We speculate that either
the tails have used up their gas, or, because the gas is not
collisionless, most of the remaining gas was largely dispersed during
the interaction.  It is worth noting, however, that clusters with age
$\tau < 10$ Myr have been reported in the tidal tails of the
Tadpole galaxy and NGC 6872 (Tran et al. 2003; Bastian et
al. 2005).  In the case of NGC 6872, the time since tail formation is
$\sim 145$ Myr (Horellou \& Koribalski 2003), making NGC 6872
significantly younger than any interacting system in our sample
(except for the pie wedge).  It is possible that NGC 6872 has simply
not yet dispersed its tidal tail gas, resulting in newly formed clusters.

Based on our cluster age distributions, we suggest the following
scenario: initially, the progenitor galaxies form clusters at a lower,
approximately constant rate. Once the two galaxies begin their
interaction, many of these disk clusters, along with other disk
material, are stripped and form the tidal tails.  The stripped gas
typically begins forming clusters immediately, resulting
in an increased rate of cluster formation.  The process continues in the
tails at a lower rate until the gas reservoir is exhausted or
the gas is dispersed.

\subsection{Comparing Cluster Ages Across Galaxies}

We suggest the following sequence of merger ages: NGC 2623 is the most
recent of the three mergers, followed by NGC 520, with NGC 3256 being
the oldest.  This sequence is supported by both our estimated cluster
ages and simulated ages of tidal tails (and thus ages of the mergers
themselves).  We note that the age sequence suggested by these simulations is
different than what the Toomre Sequence suggests, as NGC 520 and NGC
2623 are flipped (Toomre 1977).

We find that clusters in the pie wedge have ages around $\tau_{\rm{clus}} \sim 100$ Myr.
The north tail of NGC 2623 contains clusters with ages at
$\tau_{\rm{clus}} \sim 230$ Myr, while the few south tail clusters are slightly
younger, although we do not attempt to quantify their ages.  NGC 520,
the next youngest merger, exhibits a slight age
difference among clusters in different tails.  NGC 520N has clusters
around $\tau_{\rm{clus}} \sim 100$ Myr, while NGC 520S clusters has
$\tau_{\rm{clus}} \sim 260$ Myr.  The oldest merger, NGC 3256,
contains clusters in the eastern tail with ages $\tau_{\rm{clus}} \sim
260$ Myr, although we caution that the broad age distribution in NGC
3256E suggests that cluster formation likely began earlier than 260
Myr ago.  For a direct comparison, all the values of $\tau_{\rm{tail}}$
and $\tau_{\rm{clus}}$ are listed in Table \ref{Tab:clusters}.

\subsection{How are the Tidal Tail Age and Luminosity Function Related?}

We have measured the LFs for all tails where
possible, and find tentative evidence that younger mergers exhibit shallower LFs,
although our sample size is small.  M11 found that a LF with $-2.5 < \alpha
< -2$ provided qualitatively acceptable fits to their statistically
determined cluster sample in
tails where the measurement was possible, although they were not able
to directly fit the LF with a power law due to the poor quality of the data.
Whitmore et al. (2014) measured the luminosity function in a variety of spiral galaxies in
search of various correlations between $\alpha$ and galactic
environment.  A weak negative correlation was found between star
formation rate (SFR) and $\alpha$ (i.e. galaxies with high SFR tend to
have flatter LFs).  For instance, the Antennae, a major merger with a
relatively high SFR, had the second lowest $\alpha$ value in their sample, at
$\alpha=-2.07 \pm 0.03$.

No measurements of the cluster LF were made in
tidal tails, however.   While $\alpha$ tends to decrease with the tail
age, all $\alpha$ values in our sample are within the range observed
in a wide variety of other environments (Whitmore et al. 2014 and
references therein).

\section{CONCLUSIONS}

We have used ACS/WFC observations from \emph{HST} to directly observe
star clusters in the tidal tails of three nearby mergers.  We draw the
following conclusions.

\begin{enumerate}

\item Every tidal tail in our sample, as well as the pie wedge,
  contains a population of star clusters.  We note that NGC 520N and
  NGC 3256E were reported to have no statistically significant cluster
  presence in M11.

\item We estimated the ages of clusters in each tidal tail and
  compared with estimated ages for the tails themselves from simulations.
  The tidal tails in NGC 2623 contain a population of clusters which
  appear to have formed during the formation of the tails.  Tails
  of NGC 520 and NGC 3256 possess several clusters that appear to
  predate the merger, in addition to a population of
  clusters that formed at, or soon after the formation of the tails. 

\item In every tidal tail in our sample, cluster formation lasted for
  several tens of millions of years, but no very recent formation (in
  the last few million years) has occurred in any of the tails.

\item Simulations place the formation of tails in the following sequence, from
  youngest to oldest: the pie wedge, NGC 2623, NGC 520, NGC 3256.
  Ages obtained from clusters generally agree with this sequence. 

\item The diffuse light in the tails of NGC 520 and NGC 2623 exhibits
  a gradient in color across the tail (as opposed to along it), which
  is likely indicative of systematic patterns in the ages of the
  diffuse stellar population. To our knowledge, such color gradients
  in the diffuse stellar light in tidal tails of these galaxies has not been previously
  reported. The gradient is loosely traced by the spatial distribution
  of young and old star clusters within the tail.  We interpret this
  gradient as a superposition of broadly distributed older stellar
  material and younger stars and clusters that formed along the center
  of the tail, where gas is densest.

\item The LFs of clusters in our tidal tail sample
  are similar to those found in a variety of galactic environments.
  We tentatively find that as the merger age increases, the
  luminosity function tends to become steeper.
\end{enumerate}

We thank George Privon for providing us with merger simulation
results.  We also thank the anonymous referee for helpful comments.
R. C. is grateful for support from NSF through CAREER award 0847467.
This work is based on observations made with the NASA/ESA Hubble Space Telescope,
and obtained from the Hubble Legacy Archive, which is a collaboration
between the Space Telescope Science Institute (STScI/NASA), the Space
Telescope European Coordinating Facility (ST-ECF/ESA) and the Canadian
Astronomy Data Centre (CADC/NRC/CSA).  This work was supported in part
by NASA through grant GO-9735-01 from the
Space Telescope Science Institute, which is operated by AURA, INC,
under NASA contract NAS5-26555.  This research has made use of the
NASA/IPAC Extragalactic Database, which is operated by the Jet
Propulsion Laboratory, California Institute of Technology, under
contract with NASA.  This work makes use of the Digitized Sky Survey.
The Digitized Sky Surveys were produced at the Space Telescope Science
Institute under U.S. Government grant NAG W-2166. The images of these
surveys are based on photographic data obtained using the Oschin
Schmidt Telescope on Palomar Mountain and the UK Schmidt
Telescope. The plates were processed into the present compressed
digital form with the permission of these institutions.

\end{document}